\documentclass[12pt,preprint2]{aastex}
\usepackage{color}

\shorttitle{CAPELLA'S PHOTOSPHERIC ABUNDANCES}
\shortauthors{Takeda et al.}

\begin{document}


\title{SPECTROSCOPIC DETERMINATION OF CAPELLA'S PHOTOSPHERIC ABUNDANCES: \\
POSSIBLE INFLUENCE OF STELLAR ACTIVITY
}



\author{
{\sc Yoichi Takeda}\altaffilmark{1,2},
{\sc Osamu Hashimoto}\altaffilmark{3},
 {\sc and}
{\sc Satoshi Honda}\altaffilmark{4}
}
\altaffiltext{1}{National Astronomical Observatory, 2-21-1 Osawa, 
Mitaka, Tokyo 181-8588, Japan}
\email{takeda.yoichi@nao.ac.jp, osamu@astron.pref.gunma.jp, honda@nhao.jp}
\altaffiltext{2}{SOKENDAI, The Graduate University for Advanced Studies, 
2-21-1 Osawa, Mitaka, Tokyo 181-8588, Japan}
\altaffiltext{3}{Gunma Astronomical Observatory, 6860-86 Nakayama, 
Takayama, Agatsuma, Gunma 377-0702, Japan}
\altaffiltext{4}{Nishi-Harima Astronomical Observatory, Center for Astronomy,\\
University of Hyogo, 407-2 Nishigaichi, Sayo-cho, Sayo, Hyogo 679-5313, Japan}


\begin{abstract}
Capella is a spectroscopic binary consisting of two G-type giants, where 
the primary (G8~{\sc iii}) is a normal red-clump giant while the secondary
(G0~{\sc iii}) is a chromospherically-active fast rotator showing considerable 
overabundance of Li as Li-enhanced giants.
Recently, Takeda \& Tajitsu (2017) reported that abundance ratios of
specific light elements (e.g., [C/Fe] or [O/Fe]) in Li-rich giants of 
high activity tend to be anomalously high, which they suspected to be nothing but 
superficial caused by unusual atmospheric structure due to high activity.
Towards verifying this hypothesis, we determined the elemental abundances
of the primary and the secondary of Capella based on the disentangled spectrum 
of each component, in order to see whether any apparent disagreement exists 
between the two, which should have been formed with the same chemical composition.
We found that the primary is slightly supersolar (by $\sim +0.1$~dex) 
while the secondary is subsolar (by several tenths dex) for heavier elements 
such as Fe, resulting in a marked discrepancy between the primary and secondary, 
though such a trend is not seen for light elements (e.g., C or O). 
These observational facts suggest that anomalously large [X/Fe] ratios found
in Li-rich giants were mainly due to an apparent decrease of Fe abundance,
which we speculate is caused by the overionization effect due to 
chromospheric UV radiation.
We thus conclude that conventional model atmosphere analysis would
fail to correctly determine the abundances of fast-rotating giants of
high activity, for which proper treatment of chromospheric effect
is required for deriving true photospheric abundances.
\end{abstract}





\section{INTRODUCTION}

Capella (= $\alpha$~Aur = HD~34029 = HR~1708 = HIP~24608) is a spectroscopic
binary system (orbital period: 104~d) consisting of G8~{\sc iii} and G0~{\sc iii} 
giants, where the more evolved former/primary has a slightly larger 
mass and luminosity (2.6~$M_{\odot}$ and 79~$L_{\odot}$) than that of
the latter/secondary (2.5~$M_{\odot}$ and 73~$L_{\odot}$).
The primary is a typical late-G giant presumably in the He-burning
stage (red clump), which is Li-deficient and a slow rotator as other normal giants. 
In contrast, the secondary is a fast rotator (projected rotational velocity is 
$v_{\rm e}\sin i \sim 35$~km~s$^{-1}$) 
with high stellar activity (characterized by conspicuous chromospheric emission lines
in UV; see, e,g., Sanad 2013), and shows a remarkably strong Li line,
which indicates that the initial Li content is almost retained without being diluted
(the surface Li composition for the secondary is by $\sim 100$ times as high 
as the primary). That is, the secondary star belongs to the unusual group of 
Li-rich giants.

Recently, Takeda \& Tajitsu (2017; hereinafter referred to as Paper~I) conducted
a spectroscopic study on 20 Li-rich giants, in order to clarify their observational
characteristic in comparison with a large number of normal red giants.
According to their study, a significant fraction of Li-rich giants 
rotate rapidly and show high chromospheric activity (though less active
slow rotators similar to normal giants do exist). 
Curiously, they found that abundance ratios of specific light elements (e.g., 
[C/Fe] or [O/Fe] derived from high-excitation lines) in such Li-rich giants of 
high rotation/activity tend to be unusually high (typically by several tenths dex) 
and do not follow the mean trend exhibited by normal red giants (see Figure~14 
in Paper~I). Since the initial compositions of these stars should not be so 
anomalous (most of the sample stars were of normal thin-disk population)
and such overabundances of these elements are unlikely to be produced by dredge-up 
of nuclear-processed material, they suspected as a possibility that this anomaly 
is nothing but a superficial phenomenon (i.e., not the real abundance peculiarity) 
caused by unusual atmospheric structure due to high chromospheric activity.

It occurred to us that the Capella system is an adequate testbench
to verify this hypothesis, because the primary is a quite normal giant star
while the secondary is a Li-rich giant of high rotation and high activity. 
That is, these two components should have been formed with the same initial composition, 
and the abundances of the normal primary are expected to be reliably determined. 
Then, if we could detect any apparent discrepancy between the abundances
of the primary and the secondary, we may state that some problem exists
in the abundances derived for the secondary star, for which the standard
procedure of conventional model atmosphere analysis would not be sufficient 
any more. 

Unfortunately, despite that many papers have been published so far
regarding the orbital elements and fundamental stellar quantities of 
the Capella system (see, e.g., Torres et al. 2009, 2015, and the references 
therein), spectroscopic studies on the photospheric abundances are apparently
insufficient. Admittedly, the surface Li abundances of both components have 
already been clarified by several useful articles (e.g., Pilachowski \& Sowell 
1992, Torres et al. 2015, and the references therein), by which the distinct 
difference between them has been firmly established. However, few reliable 
studies appear to be available regarding Capella's abundances of various elements 
other than Li. Somewhat surprisingly, even the metallicity (e.g., Fe abundance)
of the sharp-lined primary star has not yet been established, for which considerably 
diversified results (from metal-poor through metal-rich) have been reported
(cf. Torres et al. 2009, 2015, and the references). Above all, it is only 
Torres et al. (2015) that has ever conducted separate determinations of 
elemental abundances for the primary and secondary components, as far as we know.

This scarcity may be closely connected with the practical difficulty involved 
with Capella (double-line spectroscopic binary with components of similar 
luminosity), in which the line strengths become apparently weaker because 
of dilution. Moreover, determining the abundances of the secondary star
is a very tough matter, because lines are appreciably wide and shallow 
due to large $v_{\rm e}\sin i$. In our opinion, precise abundance determination
is hardly possible by directly working on the complex double-line spectrum.
Therefore, the most promising way would be to use the adequately disentangled 
spectrum of each component, as recently done by Torres et al. (2015).

Motivated by this situation, we decided to carry out a detailed model-atmosphere
analysis for the primary and secondary components of Capella, closely following 
the procedure adopted in Paper~I, in order to see whether any meaningful 
difference exists in the photospheric abundances between these two stars.   
For this aim, the genuine single-line spectra for both stars were recovered 
by using the spectrum-disentangle method based on a set of original double-line 
spectra obtained at various orbital phases. The purpose of this paper is 
to report the outcome of this investigation.

The remainder of this article is organized as follows: 
We first explain our observational data in Section~2 and then 
the derivation of the disentangled spectra in Section~3.
Section~4 describes the spectroscopic determination of atmospheric parameters by 
using the equivalent widths of Fe~{\sc i} and Fe~{\sc ii} lines as done in Paper~I.
In Section~5 are described the abundance determinations for the primary and 
secondary components (along with the Sun) by using two different approaches:
spectrum-fitting method (as used in Paper~I) and the traditional method using
the equivalent widths. Section~6 is the discussion section, where the resulting 
abundances as well as their trends are reviewed and the implications from
these observational facts are discussed. 
We also argue especially about the recent similar abundance study of Capella
by Torres et al. (2015), because they reported a conflicting consequence with ours.
The conclusions are summarized in Section~7.
In addition, a supplementary section is also prepared (Appendix~A), 
where we check from various viewpoints whether our disentangled 
spectra have been correctly reproduced.  

\section{OBSERVATIONAL DATA}

Most of the observational data of Capella used for this study were obtained
by spectroscopic observations of 11 times over 4 months in the 2015--2016 season 
by using GAOES (Gunma Astronomical Observatory Echelle Spectrograph) installed 
at the Nasmyth Focus of the 1.5~m reflector of Gunma Astronomical Observatory, 
by which spectra with the resolving power of $R \sim 40000$ (corresponding 
to the slit width of 2$''$) covering the wavelength range of 5000--6800~\AA\ 
were obtained (33 echelle orders, each covering $\sim$~70--90~\AA).
In addition, since these GAOES spectra do not cover the long wavelength region, 
we also used the spectrum (5100--8800~\AA; $R\sim 100000$) obtained on 2010 April 29 
by using the 1.88~m telescope with HIDES (HIgh Dispersion Echelle Spectrograph) 
at Okayama Astrophysical Observatory, which was used only for the analysis 
of O~{\sc i} 7771--5 lines. 

The data reduction of these data (bias subtraction, flat-fielding, 
aperture-determination, scattered-light subtraction, 
spectrum extraction, wavelength calibration, and continuum-normalization) 
was performed using the ``echelle'' package of IRAF.\footnote{IRAF is 
distributed by the National Optical Astronomy Observatories, 
which is operated by the Association of Universities for  Research  
in Astronomy, Inc., under cooperative agreement with the National 
Science Foundation.} The S/N ratios of the resulting spectra turned out
sufficiently high (on the order of several hundred) in all cases.
The Julian dates as well as the corresponding orbital phases of 
these spectra are summarized in Table~1.

\section{SPECTRUM DISENTANGLING}

For the purpose of obtaining the pure spectrum of each star to be used for our analysis, 
we made use of the public-domain software CRES\footnote{http://sail.zpf.fer.hr/cres/} 
written by S. Iliji\'{c}. This program extracts the pure single-line spectrum of 
each component based on a set of observed double-line spectra at various orbital 
phases by the spectrum disentangling technique formulated in the wavelength domain,
which is computationally more time-consuming but more flexible than 
the one formulated in the Fourier domain (see, e.g., Simon \& Sturm 1994; 
Hadrava 1995; Iliji\'{c} 2004; Hensberge et al. 2008).

It is necessary to know the radial velocities (local topocentric radial 
velocity; $V^{\rm local}$) of both the primary and secondary components 
for each spectrum  as the input data to CRES. 
While $V_{\rm p}^{\rm local}$ for the sharp-lined primary can easily be determined,
it is difficult to precisely establish $V_{\rm s}^{\rm local}$ for the secondary
by direct measurement in the complex double-line spectrum, because lines are broad 
and shallow. Therefore, we adopted the following procedure.
We first converted $V_{\rm p}^{\rm local}$ to $V_{\rm p}^{\rm hel}$  (heliocentric
radial velocity) by applying the heliocentric correction ($\delta V^{\rm hel}$),
and plotted such derived $V_{\rm p}^{\rm hel}$ against the orbital phase ($\phi$). 
Here, $\phi$ for each spectrum was computed by using the ephemeris of
HJD($\phi = 0$) = 2450857.21 + 104.02173$E$ ($E$: integer), where the origin
as well as the orbital period were taken from Table~2 (CfA observation) and Table~9
of Torres et al. (2009).  According to their best orbital solutions
(where the velocity amplitudes are nearly the same and the eccentricity is almost zero),
the $\phi$-dependences of $V_{\rm p}^{\rm hel}$ and $V_{\rm s}^{\rm hel}$ are expressed as
\begin{equation}
V_{\rm p}^{\rm hel} = +25.96(\pm 0.04) \cos(\phi) + 29.66 (\pm 0.04) \;\; ({\rm km}\;{\rm s}^{-1})
\end{equation}
and
\begin{equation}
V_{\rm s}^{\rm hel} = -26.27(\pm 0.09) \cos(\phi) + 29.66 (\pm 0.04) \;\; ({\rm km}\;{\rm s}^{-1}).
\end{equation}
However, we found that our $V_{\rm p}^{\rm hel}$ values did not necessarily match Equation~(1) 
(cf. the dashed line in Figure~1). Therefore, we decided to slightly adjust the system velocity 
(by $-1.7$~km~s$^{-1}$), and adopt the relation (with rounded coefficients)  
\begin{equation}
V_{\rm p}^{\rm hel} = +26 \cos(\phi) + 28 \;\; ({\rm km}\;{\rm s}^{-1}),
\end{equation}
for the primary, and its reflectionally symmetric curve for the secondary
\begin{equation}
V_{\rm s}^{\rm hel} = -26 \cos(\phi) + 28 \;\; ({\rm km}\;{\rm s}^{-1})
\end{equation}
(cf. the solid lines in Figure~1).
Further, making use of these symmetric relations, we simply put $V_{\rm s}^{\rm hel}$ 
as being equal to $56 - V_{\rm p}^{\rm hel}$~(km~s$^{-1}$) at each phase $\phi$,
from which $V_{\rm s}^{\rm local}$ can be inversely evaluated by applying $\delta V^{\rm hel}$.
We eventually confirmed that such evaluated $V_{\rm s}^{\rm local}$'s in combination
with the measured $V_{\rm p}^{\rm local}$'s surely resulted in successful disentangling. 
(See also Section A.2 of Appendix~A.)
These radial velocities corresponding to each spectrum are summarized
in Table~1.

In addition, we need to specify the flux ratio ($f_{\rm p} : f_{\rm s}$), 
which is wavelength dependent. Consulting Figure~1 of Torres et al. (2015), 
we assume $f_{\rm p}/f_{\rm s}$ = 0.84, 0.97, and 1.04 at 5000, 6000, and 7000~\AA,
respectively. The relevant $f_{\rm p}/f_{\rm s}$ to be passed to CRES
was evaluated at the middle wavelength of the spectrum for each of the 33 echelle 
orders by applying quadratic interpolation.

The basic dataset is the 11 continuum-normalized GAOES spectra ($r^{\rm obs}$) 
observed at various phases (Table~1), which are overplotted in Figure~2a
(for the 5900--6080~\AA\ region).
Applying CRES to these spectra along with the relevant values of 
$V_{\rm p}^{\rm local}$, $V_{\rm s}^{\rm local}$, and $f_{\rm p}/f_{\rm s}$,
we could obtain the disentangled spectrum for the primary ($r_{\rm p}^{\rm cres}$)
and secondary ($r_{\rm s}^{\rm cres}$).
As to the sampling, we adopted 0.025\AA\ (almost equal to the original sampling)
for the primary, while somewhat larger 0.05\AA\ was adopted for the secondary. 
Note that the resulting $r_{\rm p}^{\rm cres}$ and $r_{\rm s}^{\rm cres}$
are not adequately normalized any more, as depicted in Figure~2b.
This is because what is guaranteed by CRES is that the relation 
\begin{equation}
r^{\rm obs} = f_{\rm p} r_{\rm p}^{\rm cres}({\rm shifted}) +
f_{\rm s} r_{\rm s}^{\rm cres}({\rm shifted}) 
\end{equation}
($f_{\rm p} + f_{\rm s} = 1$) holds, and thus ambiguities still exist 
in the absolute values of $r_{\rm p}^{\rm cres}$ and $r_{\rm s}^{\rm cres}$.
Therefore, an appropriate offset correction has to be applied.
For this purpose, we defined the apparent continuum (envelope) for
$r_{\rm p}^{\rm cres}$ and $r_{\rm s}^{\rm cres}$, which we denote
$c_{\rm p}$ and $c_{\rm s}$, respectively.\footnote{
Note that $c_{\rm p}$ and $c_{\rm s}$ are symmetric to each other around 
the level of unity (cf. Figure~2b), which is explained as follows: 
Because of the near equality  $f_{\rm p} \simeq f_{\rm s}$ in the present case, 
Equation~(3) reduces to $r^{\rm obs} \simeq 
0.5(r_{\rm p}^{\rm cres} + r_{\rm s}^{\rm cres})$, which means that the relation
$1 \simeq 0.5(c_{\rm p}+c_{\rm s}$) holds for the continuum level.}
Since this envelope level ($c$) should have been unity, we corrected 
$r^{\rm cres}$ to obtain the finally adopted disentangled spectrum 
($r_{\rm p}^{\rm adopt}$ and $r_{\rm s}^{\rm adopt}$) simply as
\begin{equation}
r_{\rm p}^{\rm adopt} = (1 - c_{\rm p}) + r_{\rm p}^{\rm cres}
\end{equation} 
and 
\begin{equation}
r_{\rm s}^{\rm adopt} = (1 - c_{\rm s}) + r_{\rm s}^{\rm cres},
\end{equation} 
as depicted in Figure~2c and Figure~2d, respectively.

We also checked whether the resulting $r_{\rm p}^{\rm adopt}$ and 
$r_{\rm s}^{\rm adopt}$ adequately reproduce the original spectrum 
($r^{\rm obs}$) when added with the weights of $f_{\rm p}$ and $f_{\rm s}$ 
and shifted by $V_{\rm p}^{\rm local}$ and $V_{\rm s}^{\rm local}$.
Such an example is demonstrated in Figure~2e (6015--6025~\AA\ region; 
$\phi = 0.919$), where we can see that a satisfactory match is accomplished.
(See also Section A.1 of Appendix~A.)

Although we confirmed that such obtained disentangled spectra are sufficiently 
clean and satisfactory in most of the wavelength regions, this procedure does 
not work well in specific spectral regions where strong telluric lines exist.
Therefore, regarding the 6290--6310\AA\ region comprising the important 
[O~{\sc i}] 6300 line, which is used for the spectrum-synthesis analysis, 
we specially removed the telluric line in the original $r^{\rm obs}$ by 
dividing by the spectrum of a rapid rotator before applying CRES, which 
turned out successfully. The complete set of finally adopted disentangled 
spectra ($r^{\rm adopt}$; 33 orders totally covering 4940--6790\AA, plus 
special telluric-removed spectrum in the 6290--6310\AA\ region) for 
the primary and secondary, which we used for this study, are presented 
as the online material (``spectra\_p.txt'' and ``spectra\_s.txt'').

\section{ATMOSPHERIC PARAMETERS}

Our determination of atmospheric parameters [effective temperature 
($T_{\rm eff}$), surface gravity ($\log g$), microturbulence ($v_{\rm t}$), 
and metallicity ([Fe/H])] for the primary and the secondary was implemented 
by using the TGVIT program (Takeda et al. 2005b) in the same manner as described 
in Takeda et al. (2008; see Section~3.1 therein for the details) based on 
the equivalent widths ($W_{\lambda}$) of Fe~{\sc i} and Fe~{\sc ii} lines 
measured on the disentangled spectrum of each star.
See Table~E1 of Takeda et al. (2005b) for the list of Fe lines
and their atomic data.

Regarding the sharp-lined primary star, the strengths of these Fe lines were 
measured by the Gaussian-fitting method, which was rather easy and worked quite well.   
However, measuring the equivalent widths of secondary star was a very difficult task,
because almost all lines (considerably broadened due to large rotational velocity) 
more or less suffer contamination with other lines, and thus very few blend-free lines 
are available. Therefore, special attention had to be paid as described below:
\begin{itemize}
\item
A specifically devised function was used for the fitting, which was constructed by
convolving the rotational broadening function with the Gaussian function
in an appropriately adjusted proportion.
\item
In order to decide whether a line is to be measured or not, we compared the stellar 
spectrum with the solar spectrum as well as with the theoretically computed strengths
of the neighborhood lines by using the line list of Kurucz \& Bell (1995).
\item
Judgement was done based on the following three criteria: (1) Does the wavelength 
at the flux minimum (i.e., line center) coincide with the line wavelength?
(2) Is the line shape clearly defined as expected (whichever partial or total)?
(3) Can we expect that the blending effect is not very serious (even though
influence of contamination is more or less unavoidable)? 
\end{itemize}    
The actual examples of how we measured the equivalent widths are shown
for the representative six lines in Figure~3, where we can see that
rather tricky measurement had to be done in some cases of secondary stars 
(e.g., Figure~3e; which is the case judged to be barely measurable).
In applying TGVIT, we restricted to using lines weaker than 120~m\AA\
and lines yielding abundances of large deviation from the mean ($> 2.5\sigma$)
were rejected in the iteration cycle, as done in Takeda et al. (2008).
   
The resulting parameters ($T_{\rm eff}$, $\log g$, $v_{\rm t}$) are
$4943 \pm 23$~K, $2.52 \pm 0.08$~dex, $1.47 \pm 0.13$~km~s$^{-1}$
for the primary (from  147 Fe~{\sc i} and 16 Fe~{\sc ii} lines)
and 
$5694 \pm 73$~K, $2.88 \pm 0.17$~dex, $2.29 \pm 0.38$~km~s$^{-1}$
for the secondary (from  52 Fe~{\sc i} and 6 Fe~{\sc ii} lines),
where the quoted errors are internal statistical errors
(see Section~5.2 in Takeda et al. 2002).
The mean Fe abundance ($\langle A({\rm Fe})\rangle$)\footnote{
We express the logarithmic number abundance of an element X by $A$(X), 
which is defined in the usual normalization of $A$(H) = 12.00;
i.e., $A({\rm X}) = \log [N({\rm X})/N({\rm H})] + 12.00$.}
 derived as a by-product (from Fe~{\sc i} lines) is
$7.60 \pm 0.05$ ([Fe/H] = $+0.10$) for the primary and
$7.42 \pm 0.08$ ([Fe/H] = $-0.08$) for the secondary,
where the error is the mean error ($\equiv \sigma/\sqrt{N}$, where
$\sigma$ is the standard deviation and $N$ is the number of lines).
Note that, since the $gf$ values of the adopted lines are ``solar $gf$ values''
obtained by assuming $A_{\odot}$(Fe) = 7.50 (Takeda et al. 2005b), we can 
obtain the metallicity (Fe abundance relative to the Sun) as 
[Fe/H] = $A$(Fe) $-$ 7.50.

The $A$(Fe) values corresponding to the final solutions are plotted against 
$W_{\lambda}$ (equivalent width), $\chi_{\rm low}$ (lower excitation potential),
and $\lambda$ (wavelength) in Figure~4, where we can see that no systematic 
trend is observed in terms of these quantities. We can thus confirm that the 
conditions required by the TGVIT program are reasonably fulfilled, which are the 
independence of $A$(Fe) upon $W_{\lambda}$ as well as $\chi_{\rm low}$ and the 
equality of the mean Fe abundances derived from Fe~{\sc i} and Fe~{\sc ii} lines. 
It is apparent from Figure~4 that the number of available lines is appreciably
smaller ($\sim 1/3$) and the dispersion of the abundances is evidently larger 
for the secondary star (upper panels) compared to the case of primary (lower panels),
which has caused the difference in the extent of parameter errors between two stars. 
This manifestly reflects the difficulty in measuring $W_{\lambda}$ of the secondary.

The detailed $W_{\lambda}$ and $A$(Fe) data for each line are given 
in ``felines.dat'' (supplementary online material).
The model atmospheres for the primary and secondary stars to be used in this 
study were generated by interpolating Kurucz's (1993) ATLAS9 model grid 
in terms of $T_{\rm eff}$, $\log g$, and [Fe/H]. 
Regarding the solar photospheric model used for determining the reference
solar abundances (cf. Section~5), we also used Kurucz's (1993) ATLAS9 model 
of the solar composition with $T_{\rm eff}$ = 5780~K, $\log g = 4.44$, 
and $v_{\rm t} = 1.0$~km~s$^{-1}$.

\section{ABUNDANCE DETERMINATION}

\subsection{Synthetic Spectrum Fitting}

Closely following Paper~I, we determined the abundances of various elements 
for the primary as well as secondary stars (and the Sun, for which the Moon 
spectra taken from the Okayama spectral database were used; cf. Takeda 2005a) 
by applying the spectrum-fitting technique and by carrying out a non-LTE analysis 
for several important elements (Li, C, O, Na, and Zn) based on the equivalent widths 
inversely determined from the best-fit abundance solutions.
The selected wavelength regions are as follows:
5378.5--5382~\AA\ (for C, Ti, Fe, Co), 
6080--6089~\AA\ (for Si, Ti, V, Fe, Co, Ni), 
6101--6105~\AA\ (for Li, Ca, Fe),
6157--6164~\AA\ (for Na, Ca, Fe, Ni),
6297--6303~\AA\ (for O, Si, Sc, Fe),
6361--6365~\AA\ (for O, Cr, Fe, Ni, Zn),
6704--6711~\AA\ (for Li, Fe),  
6751--6760~\AA\ (for S, Fe), and 
7768--7786~\AA\ (for O, Fe).
See Paper~I (Section~7 and Section~9 therein) for more detailed explanations 
of the procedures and the atomic data, which we adopted in this study almost unchanged.
Regarding the macroscopic broadening function to be convolved with the
intrinsic line profile, we adopted the rotational broadening function
for the secondary star, while the Gaussian function was used for the primary
star and the Sun. 
The different points compared to the case of Paper~I are as follows:\\
--- In the fitting involving the S~{\sc i} 6757 line, the wavelength range was 
made wider (6751--6760\AA) to include the Fe~{\sc i} lines at $\sim$6752--6754\AA, 
while $\sim$6754--6756\AA\ region (where observed features can not be well 
reproduced by theoretical calculations) was masked.\\
--- In the fitting involving O~{\sc i} 7771--5 triplet lines, we had to use the 
double-line spectrum (at $\phi = 0.863$) observed by OAO/HIDES, because
this triplet is outside of the wavelength range of the GAOES (i.e., disentangled) spectra.
For this purpose, we modified our automatic solution-search program 
based on the algorithm described in Takeda (1995a), so that it may be applied
to the combined spectra of primary and secondary components.\footnote{
In this case, the problem becomes much more complicated and difficult since 
the total number of the parameters (elemental abundances, macrobroadening
parameters, radial velocity, etc.) to be determined is almost doubled.
Actually, we did not vary the whole set of the parameters 
($\alpha_{1}^{\rm p}$, $\alpha_{2}^{\rm p}$, $\ldots$ $\alpha_{M}^{\rm p}$
for the primary; 
$\alpha_{1}^{\rm s}$, $\alpha_{2}^{\rm s}$, $\ldots$ $\alpha_{N}^{\rm s}$
for the secondary) simultaneously. Instead, we first varied only $\alpha^{\rm p}$'s for the primary
and had them converged, while $\alpha^{\rm s}$'s for the secondary being fixed. 
Then $\alpha^{\rm s}$'s for the secondary were varied and made converged, 
while $\alpha^{\rm p}$'s for the primary being fixed. And these two consecutive runs were 
repeated several times until satisfactory convergence for all parameters has been accomplished.
We confirmed that this procedure worked well, although the starting parameters 
had to be carefully adjusted so as to be sufficiently close to the final solutions.
} 
Also, the wavelength range was chosen wider (7768--7786~\AA) in order to
cover the two spectra of considerable radial velocity difference.

How the fitted theoretical spectrum matches the observed spectrum (for the secondary, 
the primary, and the Sun) is displayed for each region in figure~5.
The equivalent widths inversely derived from the abundance solutions and the 
non-LTE corrections for the important lines (C~{\sc i}~5380, Li~{\sc i}~6104,
Na~{\sc i}~6161, Zn~{\sc i}~6362, S~{\sc i}~6757, and O~{\sc i}~7774)
are given in Table~2. 
The finally obtained fitting-based abundances relative to the Sun for the primary 
([X/H]$_{\rm p}$) as well as the secondary ([X/H]$_{\rm s}$), and the secondary$-$primary 
differential abundances [X$_{\rm s}$/X$_{\rm p}$] are presented in Table~3. 

\subsection{Abundances from Equivalent Widths}

We also carried out differential analysis of the elemental abundances (other than Fe) 
of the primary as well as the secondary relative to the Sun, based on the equivalent 
widths of usable spectral lines, which were measured on the disentangled spectra 
as done in Section~4. The procedures of this analysis are detailed in Section~4.1 
of Takeda et al. (2005c). Note that the solar equivalent widths used as the reference 
for this analysis were derived by Gaussian-fitting measurement on Kurucz et al.'s 
(1984) solar flux spectrum atlas.  Regarding the primary, we measured the
equivalent widths for 173 lines of 22 elements (C, O, Na, Al, Si, Ca, Sc, Ti, V, Cr, 
Mn, Co, Ni, Zn, Y, Zr, La, Ce, Pr, Nd, Gd, and Hf), while those of the secondary 
could be evaluated only for 27 lines of 9 elements (C, Al, Si, Ca, Sc, Ti, Cr, Ni, La),
where the measurement was done in the same manner as the case of Fe lines described 
in Section~4. The resulting differential abundances relative to the Sun
for each of the lines are separately summarized in ``ewabunds\_p.dat'' (primary) 
and ``ewabunds\_s.dat'' (secondary) of the online material.

We are afraid, however, that the number of available lines for each element
is generally not sufficiently large and the appreciable fluctuation is often seen
in the abundances, which makes the mean abundance less credible.
Given that our purpose is (not to determine the abundances of as many element 
as possible but) to establish the abundance of any element as precisely as possible, 
we decided to confine our discussion to only those elements for which 7 or more 
lines are available (i.e., Si, Ca, Sc, Ti, V, Cr, Co, and Ni for the primary, 
and only Ni for the secondary). The resulting mean differential abundances
of these elements and those of Fe (already derived in Section~4) are also 
summarized in Table~3. 

\section{DISCUSSION}

\subsection{Photospheric Abundances of Heavy Elements}

The differential abundances relative to the Sun for the secondary ([X/H]$_{\rm s}$),
those for the primary ([X/H]$_{\rm p}$), and the secondary$-$primary abundance
differences ([X/H]$_{\rm s}-$[X/H]$_{\rm p}$), which were obtained in Section~5,1
and Section~5.2 and summarized in Table~3, are plotted against the atomic number ($Z$) 
in Figure~6. Hereinafter, we distinguish the ``S''ynthesis-based abundance 
and equivalent-``W''idth based abundance by the superscript ``S'' and ``W'',
respectively, and the mean over available lines is denoted by the bracket
``$\langle$'' and ``$\rangle$''.

We first discuss the heavier elements such as those of the Fe group ($Z \ga 20$).
Let us recall that we obtained $\langle$[Fe/H]$_{\rm p}^{\rm W}\rangle = +0.10 (\pm 0.05)$ and 
$\langle$[Fe/H]$_{\rm s}^{\rm W}\rangle = -0.08 (\pm 0.08)$ from determinations 
of atmospheric parameters based on Fe~{\sc i} and Fe~{\sc ii} lines (cf. Section~4): 
That is, the primary/secondary are slightly metal-rich/-poor by $\sim 0.1$~dex, which means 
that the metallicity of the secondary is less than the primary by $\sim 0.2$~dex.

This trend is confirmed also by abundances for other $Z \ga 20$ elements shown in Figure~6.
Regarding the primary, both of the $\langle$[X/H]$_{\rm p}^{\rm S}\rangle$ 
and $\langle$[X/H]$_{\rm p}^{\rm W}\rangle$ for each element is $\sim$~+0.1--0.2~dex,
which almost agree with the case of Fe (cf. Figure~6a).    
As to the secondary, while the W-based abundance (only for Ni) is 
$\langle$[Ni/H]$_{\rm s}^{\rm W}\rangle = -0.10$ and consistent with 
$\langle$[Fe/H]$_{\rm s}^{\rm W}\rangle$, the S-based abundances 
for each element are somewhat more metal-poor to be 
$\langle$[X/H]$_{\rm s}^{\rm S}\rangle \sim -$0.2--0.4 (see Figure~6b).
In any event, we can conclude that, for heavier elements ($Z \ga 20$),
the primary is slightly metal-rich ($\langle$[X/H]$_{\rm p}\rangle \sim +0.1$ to +0.2)
while the secondary is apparently metal-poor ($\langle$[X/H]$_{\rm s}\rangle \sim -0.1$
to $-0.4$; the extent being appreciably line-dependent), which means that the secondary 
is superficially metal-poor compared to the primary by several tenth dex (cf. Figure~6c).

Regarding the primary star, its marginally metal-rich nature (by 
$\sim $0.1--0.2~dex) is reasonably consistent with the fact that Capella 
belongs to the Hyades moving group (van Bueren 1952; Eggen 1960, 1972),
because the metallicity ([Fe/H]) of this moving group is known to be
slightly supersolar\footnote{Exceptionally, Zhao et al. (2009) reported
a slightly subsolar mean metallicity of $-0.09$~dex (the standard deviation 
of $\sigma = 0.17$) for the Hyades moving group. However, details are not 
described in their paper regarding how this metallicity value was derived,
and the dispersion is apparently too large. Accordingly, their result 
seems to be less credible.} as +0.11 (Boesgaard \& Budge 1988) or +0.07 (Fuhrmann 2011), 
in agreement with that of the Hyades cluster ($\sim$+0.1--0.2~dex; see, e.g., Takeda 2008; 
Takeda et al. 2013). 

\subsection{Behaviors of Light Elements}

Interestingly, such a tendency as shown by heavier species is not observed
for lighter elements, as can be confirmed in Figure~6. 
Before discussing this trend, however, we should keep in mind that 
the surface abundance of light elements may have suffered changes due to 
a dredge-up of nuclear-processed material caused by evolution-induced mixing 
in the stellar envelope. This possibility is relevant for Li, C, O, and Na 
in the present case. For reference, we depict in Figure~7 how the surface 
abundances of these elements (main isotopes) are theoretically expected to 
undergo changes during the course of stellar evolution, which were simulated by
Lagarde et al. (2012) for the standard mixing model and the special mixing model
including rotational+thermohaline mixing.

Regarding Li, we obtained $A_{\rm p} \simeq 1.0$ (primary) and 
$A_{\rm s} \simeq 3.2$ (secondary), which are almost consistent with
the previous studies (see, e.g., Pilachowski \& Sowell 1992 or
Torres et al. 2015, and the references therein). This means that the former
is within the expected range shown by normal giants while the latter
is the typical abundance of Li-rich giants (cf. Figure~11 in Paper~I).
Let us postulate (as most previous studies argued) that the secondary is still 
on the way of ascending the red-giant branch and preserves the original Li 
composition without being diluted. Then, the following characteristics
can be read from Figure~7:\\
--- The secondary star should have rotated rather rapidly (presumably 
$v_{\rm e} \sim 100$~km~s$^{-1}$) when it was on the main sequence, 
as expected from the current $v_{\rm e}\sin i$. This is almost the same
initial rotation rate assumed by Lagarde et al. (2012; see Section~2.3 therein) 
for their rotational+thermohaline mixing model, which indicates a significant 
dilution of Li already just after evolving off main-sequence stage 
(cf. solid line in Figure~7a). Then, if the secondary suffered no dilution, 
we may state that Lagarde et al.'s (2012) rotational+thermohaline mixing model 
is not realistic (which overestimates the extent of mixing) but the standard 
mixing may be more adequate.\\
--- Li is the most fragile species (which is destroyed by reaction with 
proton at a comparatively low temperature of $T \sim 2.5 \times 10^{6}$~K) 
among all elements, and its surface abundance is quite vulnerable to mixing
in the envelope (cf. Figure~7a). So, if Li is preserved in the surface of 
the secondary star, any other elements must retain their original composition 
at the time of star formation. 
We may thus state that the primordial abundances of C, O, and Na are 
kept unchanged in the photosphere of secondary star.\\
---  On the other hand, we can reasonably assume that the primary star is 
a normal red-clump giant in the core He-burning stage, such as expected for 
many red giants studied by Takeda et al. (2015). According to their observational
results (see Fig.~10 therein), we may regard that C is moderately deficient 
by $\sim 0.2$~dex, O is only slightly deficient by $\la 0.1$~dex, 
and Na is mildly overabundant by $\sim 0.2$~dex, which are quite consistent 
with the predictions of theoretical simulations (cf. Figure~7).

Keeping this in mind, we can draw the following consequences by comparing 
the abundances of these light elements for the primary and the secondary stars:\\
--- (1) Regarding carbon, the results of [C/H]$_{\rm p}\sim -0.2$ and 
[C/H]$_{\rm s}\sim 0.0$ are just consistent with the expectation mentioned above.
This means that the correct photospheric C abundances could be derived
from the C~{\sc i} 5380 line for both components.\\
--- (2) As to the case of oxygen, only  O~{\sc i} 7771--5 triplet lines are usable 
for comparing the O-abundances of both components. The abundances derived from this triplet 
are [O/H]$_{\rm p} = +0.08$ and [O/H]$_{\rm s} =+0.21$ for the primary and the secondary, 
respectively; the latter being slightly larger (by $\sim 0.1$~dex) than the former. 
Considering that the surface O abundance for the primary may have suffered only a 
slight decrease ($\la 0.1$~dex) due to envelope mixing while the secondary should 
retain the original composition, we can recognize a reasonable consistency
between the oxygen abundances of the primary and the secondary; its original 
composition would have been slightly supersolar ([O/H]~$\sim$~0.1--0.2)
such as the case of [Fe/H]$_{\rm p}$ (indicating [O/Fe]~$\sim 0$).\\
--- (3) The sodium abundances derived from the Na~{\sc i} 6161 line are of particular 
interest, which turned out [Na/H]$_{\rm p} = +0.41$ and [Na/H]$_{\rm s} = -0.22$ 
for the primary and the secondary, respectively. Considering that the primary 
has experienced an enrichment of surface Na by $\sim +0.2$~dex due to dredge-up of 
nuclear-processed product as mentioned above, we may state that its primordial Na 
abundance should have been slightly supersolar by $\sim +0.2$~dex, which is consistent 
with other heavier elements such as Fe. On the other hand, the mildly subsolar tendency 
for the secondary ([Na/H]$_{\rm s} \sim -0.2$) is hard to be reasonably explained.
We have no other way than to consider that the sodium abundance for the secondary
was erroneously underestimated by several tenths dex, presumably because 
the conventional model atmosphere analysis is not properly applicable
any more to the secondary star. 

\subsection{Origin of Superficial Abundance Anomaly}

Based on the observational results described in Section~6.1 and Section~6.2,
we can judge whether or not our adopted standard method of analysis using LTE 
model atmospheres can yield the correct abundances of both components, 
which are summarized as follows.
\begin{itemize}
\item
Regarding the primary (normal red giant of slow rotation and low activity), the conventional 
method of abundance determination turned out to be successfully applicable without 
any problem, by which the expected abundances could be reasonably derived. 
\item
As to the secondary (rotating giant star of high chromospheric activity), apparently low 
abundances (erroneously low by several tenths dex) were derived for heavier elements 
($Z \ga 20$; such as those of Fe group) as well as for Na, though the extent of this apparent 
underabundance appears to appreciably depend on the used lines. That is,
in order to obtain the correct photospheric abundances of the secondary star
for these elements, some non-canonical method (e.g., by properly taking into account 
the chromospheric effect) has to be applied.  
\item
However, reasonably correct abundances could be obtained even for the secondary star regarding C 
(from C~{\sc i} 5380) and O (from O~{\sc i} 7771--5) by our conventional non-LTE analysis. 
\end{itemize}

Turning back to the question raised in Paper~I (i.e., regarding the anomalously large
abundance ratios such as [C/Fe] or [O/Fe] observed in Li-rich giants of high rotation and
high activity) which motivated this investigation, we have arrived at a satisfactory solution:
It was not the increase of the numerator (C or O) but the decrease of denominator (Fe) 
that caused the apparently peculiar abundance ratio. That is, Fe abundances of active Li-rich 
giants would have been erroneously underestimated.

Then, which mechanism is involved with this superficial abundance anomaly?
How could the active chromosphere affect the line formation?
In Paper~I, we once speculated that the chromospheric temperature rise in the upper atmosphere 
might have strengthen the temperature-sensitive high-excitation lines of C or O, since 
such an intensification is actually expected for O~{\sc i} 7771--5 (e.g., Takeda 1995b). 
However, this should not be the case, because correct abundances were obtained for these
light elements as mentioned above.
We have to find the reason why unusually low abundances were obtained for Fe group elements 
as well as for Na under the condition of high chromospheric activity. 
In our opinion, the most promising mechanism may be the overionization caused by excessive
UV radiation radiated back from the chromosphere. As explained in Takeda (2008), it is 
lines of minor population species such as Fe~{\sc i} (while the parent ionization stage is 
the major population such as Fe~{\sc ii}) that can be significantly weakened by overionization.
From this point of view, those elements showing appreciable underabundances in the secondary star 
(Na and heavier $Z \ga 20$ elements) are mostly derived from lines of neutral species of
minor population with comparatively low ionization potential 
(this effect is irrelevant for C~{\sc i} 5380 and O~{\sc i} 7771--5 lines, because neutral 
stage is the dominant population for C as well as O due to their high ionization potential).

However, even if this scenario is relevant, it is not easy to evaluate how much correction
is required to recover the true abundance, because it appears to considerably differ
from line to line. While only the slightly subsolar value ([Fe/H]$_{\rm s}^{\rm W} \simeq -0.1$) 
was derived from equivalent widths of weak to mildly strong lines 
($W_{\lambda} \le 120$~m\AA; cf. Section~4), 
those derived from spectrum synthesis ([Fe/H]$_{\rm s}^{\rm P}$) were more metal-poor 
(even down to $\sim -0.5$ or lower) and diversified, which may be attributed to the fact
that various lines (including strong saturated lines which form high in the atmosphere) 
are involved in the spectrum synthesis analysis. 
Besides, if an overionization effect by chromospheric radiation is important,
it is expected to have an appreciable impact on  Li lines (Li~{\sc i} 6708 and Li~{\sc i} 6104),
because neutral Li is a minor population species (i.e., mostly ionized) due to its low 
ionization potential. If so, the true Li abundance of the secondary might be significantly higher
than the value we derived (3.2 from Li~{\sc i} 6708), which makes us wonder whether it is consistent
with the assumption that the primordial Li is retained in the secondary (i.e., some further 
Li-enhancement mechanism might be responsible?).  
In any event, much more things still remain to be clarified regarding spectroscopic 
abundance determination for giant stars of high chromospheric activity. 

\subsection{Adequacy Check of Our Analysis}

\subsubsection{Conflict with Torres et al. (2015)}

As mentioned in Section~1, only a small number of spectropic studies have 
been published for Capella so far, as briefly summarized in Table~4
(see also Footnote~9 in Torres et al. 2009). 
Especially, to our knowledge, it is only the recent investigation of 
Torres et al. (2015) who carried out abundance determinations for a number 
of elements separately for the primary and secondary components, 
as done in this study. Therefore, it is worth discussing their results
in detail in comparison with our conclusion.

Torres et al. (2015) determined the abundances for 22 elements
for both components (and oxygen only for the primary) from the equivalent 
widths measured for many spectral lines (in 4600--6750~\AA) 
on the disentangled spectra (which were processed in a manner similar to ours 
based on the 15 spectra taken from the ELODIE archive; cf. Moultaka et al. 2004).
We recognize, however, a distinct discrepancy between their work and this study.
They concluded that the abundances of most elements (except for Li) are nearly solar 
(within $\sim \pm 0.1$~dex) for both components, which means that no meaningful 
compositional difference was observed between the primary and the secondary 
(cf. their Figure~3). Thus, their conclusion is in serious conflict with 
our consequence. 

Unfortunately, since Torres et al. (2015) did not publish any detailed data
(e.g., equivalent widths, $gf$ values, and abundances for individual lines 
they used), it is impossible for us to verify their results.
We would here point out two concerns regarding their analysis:\\
--- First, the number of lines
they used for abundance determination of each element is almost the same
for the primary and the secondary (see their Table~2). This is hard to 
understand, because the number of measurable lines should be considerably 
smaller for the broad-lined secondary than for the sharp-lined primary
(in our case, the number of lines usable for the secondary was about
$\sim 1/3$ of that for the primary).\\ 
--- Second, their way of deriving [X/H] (abundance relative to the Sun)
was simply to subtract the reference solar abundance (taken from Asplund et al. 2009)
from the absolute mean abundance (obtained by averaging the abundances
from individual lines); i.e., their analysis is not a differential one,
and thus directly suffer errors involved in $gf$ values. 

\subsubsection{Spectra and equivalent widths}

In any event, given that such an conflicting result (near-solar abundances for both 
components) was once reported, we had better check whether something is different or 
wrong with our data and analysis, in comparison with that of Torres et al. (2015).
Regarding the observational spectra of Capella, we can not see any significant
difference between theirs and ours, not only for the original spectra (ELODIE
vs. GAOES; cf. Figure~8a) but also for the disentangled spectra (Figure~8b,
which appears almost consitent with Figure~2\footnote{
Note that the disentangled 
spectra depicted in Figure~2 of Torres et al. (2015) appear to be normalized
with respect to the composite (primary+secondary) continuum flux (not to the 
continuum of either primary or secondary as done by us), because the line-depths 
of very strong lines do not exceed $\sim$40--50\%.
} of Torres et al. 2015).

There is a good reason to believe that our abundances for the sharp-lined primary 
are reliable because equivalent-width measurements are easy (see also Appendix~A.3).
However, considerable difficulties are involved for the broad-lined secondary
as mentioned in Section~4. Is it possible that our $W_{\lambda}$ values 
(and thus the abundances) for the secondary might have been significantly underestimated; 
e.g., by incorrect placement of the continuum level?   
We checked this point by a simple test. If the continuum level is shifted by 
$\epsilon (\ll 1)$ as $f_{\rm c} \rightarrow f_{\rm c}(1 + \epsilon)$,
the line depth $R_{\lambda} (\equiv 1 - f_{\lambda}/f_{\rm c})$ is
changed as $R_{\lambda} \rightarrow R_{\lambda} [1 + \epsilon (1- R_{\lambda})/R_{\lambda}]$.
If we presume that the equivalent width of a line ($W_{\lambda}$) is proportional to its 
line-center depth ($R_{0}$), the expected change of $W_{\lambda}$ may be expressed as 
$W_{\lambda} \rightarrow W_{\lambda} [1 + \epsilon (1- R_{0})/R_{0}]$. 
Assuming $\epsilon = 0.01$ (i.e., raising the continuum by 1\%), 
we examined how much abundance changes ($\Delta A$) would result 
for each line by this slight raise of the continuum, which are depicted in Figure~9a.
We can see from this figure that only a slight change of the continuum level by 1\% 
leads to a significant abundance variation by $\sim$~0.2--0.3~dex for the secondary,
while the abundance changes for the primary are insignificant (only a few hundredths
dex for most cases) except for very weak lines.
Therefore, we should keep in mind a possibility of appreciable systematic errors
for the abundances of the secondary star if our continuum placement was not adequate, 
as far as abundances based on equivalent widths are concerned
(i.e., those described in Section~4 and Section~5.2).
Yet, we consider such a case rather unlikely, because the secondary's abundances derived 
by using Takeda's (1995a) spectrum-fittng method (Section~5.1), which is irrelevant to 
the continuum position, similarly turned out to be significantly subsolar. 

\subsubsection{Microturbulence}

Regarding the atmospheric parameters, Torres et al.'s (2015) $T_{\rm eff}$ and 
$v_{\rm t}$ values were determined spectroscopically by using iron lines 
in a similar manner to ours, while they seem to have adopted the directly 
evaluated values for $\log g$. While we can see a reasonable consistency  
for $T_{\rm eff}$ and $\log g$ (cf. Table~4), an appreciable difference is observed for 
$v_{\rm t}$ of the secondary (though $v_{\rm t}$ for the primary is in agreement): 
that is, $1.55\pm 0.11$~km~s$^{-1}$ (theirs) and $2.29\pm 0.38$~km~s$^{-1}$ (ours).
Actually, this difference of $\sim$~0.7~km~s$^{-1}$ is so large as to
result in significant abundance changes, especially for strong saturated lines. 
In order to demonstrate this fact, we calculated the abundance differences for the
secondary star ($\delta A_{\rm s}$) caused by using Torres et al.'s (2015) 
$v_{\rm t} =1.55$~km~s$^{-1}$ instead of our 2.29~km~s$^{-1}$, which are plotted
against $W_{\lambda}$ (Figure~9b) and $Z$ (Figure~9c).  
We can see from Figure~9b that significant $W_{\lambda}$-dependent abundance variations 
(typically several tenths dex, even up to $\sim 0.6$~dex) are caused by this deacrease 
in $v_{\rm t}$.  Moreover, according to Figure~9c, appreciably large increases of
the abundances are typically seen for comparatively heavier elements ($Z>20$ species
such as Fe group), while lighter elements ($Z \le 20$) are not much affected because
lines are not so strong. Combining Figure~6b with Figure~9c, we can see that
the use of $v_{\rm t} =1.55$~km~s$^{-1}$ would bring the secondary's abundances 
nearer to the solar composition for most elements, except for the oxygen abundance
derived from the O~{\sc i} 7774 line, which would become anomalously supersolar.
Accordingly, it is likely that the difference in the adopted $v_{\rm t}$ is
the main cause for the discrepancy in the secondary's chemical composition 
between Torres et al. (2015) and this study.

Then, which of 2.29~km~s$^{-1}$ and 1.55~km~s$^{-1}$ is the correct $v_{\rm t}$ solution 
for the secondary? Considering the appreciably large abundance change ($\sim 0.4$~dex at 
$W_{\lambda} = 100$~m\AA), the latter (low-scale) solution is definitely unlikely as far as 
our $W_{\lambda}$ data are concerned, because $W_{\lambda}$-independence of Fe abundances 
is successfully accomplished with the former (high-scale) solution (cf. Figure~4d).  

From a different perspective, it is worthwhile to pay attention to the relative
comparison of [$v_{\rm t}$(primary), $v_{\rm t}$(secondary)]: our analysis
yielded markedly different values [1.47~km~s$^{-1}$, 2.29~km~s$^{-1}$], while 
Torres et al. (2015) derived almost the same results [1.48~km~s$^{-1}$, 1.55~km~s$^{-1}$].
Regarding evolved FGK giants, the behavior of microturbulence ($v_{\rm t}$) across the HR 
diagram (especially its $T_{\rm eff}$-dependence) was historically somewhat controversial.
In the review paper of Gray (1978) was suggested a tendency of increasing $v_{\rm t}$ 
with $T_{\rm eff}$ for giants and supergiants (cf. Fig.~11 therein) based on 
the results of several studies published mainly in 1970s.
Meanwhile, Gray (1982; cf. Fig.~11 therein) reported based on his line-profile analysis 
that $v_{\rm t}$ values of 4 late-type giants are almost $T_{\rm eff}$-independent at 
$\sim 1.5$~km~s$^{-1}$. See also Fig.~3-8 in the lecture book of Gray (1988). 
However, considering that the behavior of``macro''-turbulence in FGK giants has been 
almost established to show an increasing tendency not only toward higher luminosity class 
but also toward earlier spectral types (see Fig.~17.10 of Gray 2005), we may expect 
similar trend to hold also for ``micro''-turbulence.
Actually, Takeda et al. (2008) showed that $v_{\rm t}$ values of GK giants
(which were determined with the same procedure and line list as used in this study)
tend to increase with $T_{\rm eff}$ at $T_{\rm eff} \ga 5000$~K (cf. Fig.~1d therein).
Besides, this trend naturally links to the distrbution of $v_{\rm t}$ for evolved 
A-, F-, and G-type stars recently determined by Takeda, Jeong, \& Han (2018) 
from O~{\sc i}~7771--5 lines. This situation is illustrated in Figure~10, where
our $v_{\rm t}$ results for the primary and secondary are also plotted by crosses. 
This figure reveals that these crosses reasonably follow the general trend,
and thus $v_{\rm t}$ being larger for the secondary is quite natural.
Consequently, our choice of high-scale $v_{\rm t}$ (2.29~km~s$^{-1}$) for the secondary 
is considered to be justified, which may lend support to our conclusion.

\section{CONCLUSION}

Capella is a spectroscopic binary, which consists of two G-type giants with
similar mass and luminosity. An interesting feature of this system is that,
while the slightly more evolved primary (G8~{\sc iii}) is a slowly-rotating 
normal red-clump giant, the secondary (G0~{\sc iii}) is a chromospherically-active 
fast rotator ascending the giant branch and shows a marked overabundance of Li 
(i.e., Li-rich giant).

Recently, it was reported in Paper~I that abundance ratios of specific 
light elements (e.g., [C/Fe] or [O/Fe]) in Li-rich giants of high activity 
tend to be anomalously high as compared to normal giants, which they suspected 
to be nothing but a superficial phenomenon (i.e., not real) caused by unusual 
atmospheric structure due to high chromospheric activity. The Capella system is 
a suitable testbench to verify this hypothesis; that is, if we could detect any 
apparent difference between the abundances of two stars, it may lend support 
for this interpretation, since we may postulate that both were originally born 
with the same chemical composition, 

Toward this aim of searching for any apparent disagreement between the abundances of
Capella's two components, we carried out a spectroscopic analysis to determine 
the elemental abundances of the primary and the secondary of Capella.
Since it is difficult to accomplish this task by working on the complex double-line 
spectra, we first recovered the genuine spectrum of the primary and that of the secondary
by applying the spectrum-disentangle method to a set of original spectra obtained
at various orbital phases.

The atmospheric parameters ($T_{\rm eff}$, $\log g$, and $v_{\rm t}$) of 
both stars were spectroscopically determined from the equivalent widths of Fe~{\sc i} 
and Fe~{\sc ii} lines as done in Paper~I. Based on these parameters, we constructed
the model atmospheres to be used for further abundance determination.
The abundances of various elements were derived in two ways:
(i) spectrum-fitting technique (as used in Paper~I) applied to 9 wavelength regions, 
and (ii) analysis of equivalent widths as done by Takeda et al. (2005c).

We could clarify several significant observational facts and implications from the 
resulting abundances as itemized below, while postulating that (1) the primordial composition 
of the Capella system  is the same for both components, (2) the primary star is a normal giant 
having suffered surface abundance changes due to envelope mixing in several specific 
light elements (Li, C, O, Na) and (3) the secondary star retains the original 
composition in its surface unchanged for all elements,
\begin{itemize}
\item
We confirmed that the surface Li abundances are substantially different; 
$A_{\rm p} \sim 1.0$ (primary) and $A_{\rm s} \sim 3.2$ (secondary), as already 
reported by several previous studies, indicating that the former experienced 
a significant depletion while the latter retains the original Li composition. 
\item
Regarding the heavier elements ($Z \ga 20$) such as those of the Fe group,
the abundances of the primary star turned out somewhat supersolar 
([X/H]$\sim$~+0.1--0.2), which is consistent with the expectation 
because Capella belongs to the Hyades moving group.
On the contrary, we found that the [X/H] values of the secondary are
appreciably subsolar by several tenths dex (from $\sim -0.1$ down to $\sim -0.5$ 
or even lower) and diversified, which indicates an apparent disagreement 
exists between the abundances of the primary and the secondary.
\item
However, as to the light elements, such a tendency is not seen. 
For example, we can state that reasonably correct abundances of C 
(from C~{\sc i} 5380) or O (from O~{\sc i} 7771--5) could be obtained for both 
the primary and the secondary star by our conventional non-LTE analysis. Nevertheless, 
the exceptional case was Na (from Na~{\sc i} 6161), for which we found an unusual 
deficiency by several tenths dex for the secondary, despite that a reasonable
abundance was obtained for the primary.
\item
Taking these observational facts into consideration, we think we could trace down
the reason why anomalously large abundance ratios (such as [C/Fe] or [O/Fe])
were observed in Paper~I for Li-rich giants of high rotation/activity.
That is, it was not the increase of the numerator (C or O) but the decrease of 
denominator (Fe) that caused the apparently peculiar abundance ratios. In other words, 
Fe abundances of active Li-rich giants would have been erroneously underestimated.
\item
We note that lines yielding appreciable underabundances for the secondary star
are mostly of minor-population species (e.g., neutral species such as
Na~{\sc i}, Fe~{\sc i}, etc.) with comparatively low ionization potential.
Accordingly, we suspect that the overionization caused by excessive
UV radiation radiated back from the active chromosphere is responsible for the
weakening of these lines, eventually resulting in an apparent underabundance. 
\item
To conclude, the conventional model atmosphere analysis presumably fails to 
correctly determine the abundances of fast-rotating giants of high activity.
A proper treatment of chromospheric effect would be required if true photospheric 
abundances are to be derived for such stars.
\item
To be fair, our consequence is in marked conflict with the similar abundance study on 
Capella recently carried out by Torres et al. (2015), who came to the conclusion that 
the photospheric abundances for the primary and the secondary are observed to be practically
the same at the solar composition. Although we can not clarify the cause of this 
discrepancy because Torres et al. (2015) did not publish any details of their analysis,
appreciably different values of adopted microturbulence for the secondary may be an 
important factor. Having examined this problem, however, we confirmed that our choice 
is reasonably justified, which may substantiate our conclusion. 
\end{itemize}

\acknowledgments
This study partly made use of the spectral data retrieved from the ELODIE archive at 
Observatoire de Haute-Provence (OHP).
Data reduction was in part carried out by using the common-use data analysis 
computer system at the Astronomy Data Center (ADC) of the National Astronomical 
Observatory of Japan.

\appendix

\section{VALIDITY CHECK FOR THE DISENTANGLED SPECTRA}

Since our abundance determination carried out in this paper is mostly based on
the resolved spectra of the primary and secondary, which were disentangled
with the help of the CRES program from a set of actually observed 11 spectra,
it is important to ensure that this decomposition was correctly done. 
In this supplementary section, we present some additional information 
indicating that the spectra have been properly processed, where our attention 
is paid to (i) consistency between the reconstructed and original spectra,
(ii) compatibility of the line widths ($v_{\rm e}\sin i$) for the secondary, 
and (iii) comparison of the equivalent widths of the primary with those of
Hyades giants with similar parameters. 

\subsection{Matching test of reconstructed and original spectra}

A simple and reasonable test to check whether this decomposition process has been 
adequately done is to compare the original double-line spectrum (at any phase) 
with the inversely reconstructed composite spectrum defined by the following 
relation (linear combination of the disentangled primary and secondary spectra).  
\begin{equation}
r^{\rm rec} = f_{\rm p} r_{\rm p}^{\rm adopt}(v = v_{\rm p}^{\rm local}) +
f_{\rm s} r_{\rm s}^{\rm adopt}(v = v_{\rm s}^{\rm local}). 
\end{equation}
Although this check was briefly mentioned in Section~3 for a restricted 
narrow wavelength range of 6015--6025~\AA\ (cf. Figure~2e), we here show 
this comparison for further six echelle orders from shorter to longer wavelength
ranges in Figures~11a,a$'$--11f,f$'$, where the runs of $r^{\rm rec}$, $r^{\rm obs}$, 
and their difference ($r^{\rm obs} - r^{\rm rec}$) with wavelength are depicted.
We can see from these figures that the agreement between $r^{\rm rec}$ and $r^{\rm obs}$
is satisfactory in most cases, especially for longer wavelength regions (differences 
are $\la 1\%$). 

While we notice an appreciably undulated difference (up to $\la 10\%$) in shorter 
wavelength regions (cf. Figures~11a and 11b), this is not due to the disentangling 
procedure but only a superficial phenomenon caused by different nature of continuum 
normalization. That is, in preparing $r^{\rm obs}$, we normalized the raw spectrum 
($f^{\rm obs}$) by dividing it by the continuum (envelope) determined 
in the same manner as done for $c_{\rm p}$ or $c_{\rm s}$ (cf. Section~3), 
for which we tried to express the global line-free envelope by spline functions 
using the IRAF task ``continuum.'' 
Unfortunately, since the ``true'' continuum level can not exactly be represented 
in this way, this kind of normalization is more or less imperfect. 
Especially in shorter wavelength regions where lines are crowded, the specified 
tentative continuum tends to be lower than the true continuum, making the normalized flux ($r$) 
somewhat too large. Further, the situation is comparatively worse for the case of
$f^{\rm obs} \rightarrow r^{\rm obs}$ (than that for $r^{\rm cres} \rightarrow r^{\rm adopt}$; 
cf. Section~3), because the echelle blaze function affecting $f^{\rm obs}$ is generally 
more complex than the gradual undulation shown by $r^{\rm cres}$ (cf. Figure~2b). 
This is the reason for the trend of $r^{\rm obs} > r^{\rm rec}$ observed in Figures~11a 
and 8b, though the extent of difference varies from region to region. In any event, 
since this discrepancy between $r^{\rm obs}$ and $r^{\rm rec}$ is only gradual 
and can be regarded almost constant within a line profile (covering $\la$~1--2~\AA), 
we can make them match locally well by applying an appropriate offset.\footnote{   
Yet, we must note that this local discrepancy seen in specific parts of 
short-wavelength region causes an ambiguity of zero-point level in the residual 
intensity, just like the case of scattered light. Therefore, some uncertainty 
would result when equivalent widths are measured there, especially for strong 
deep lines (though insignificant for weak shallow lines). 
However, this problem is essentially irrelevant to the entire 
results of our analysis, because we barely measured equivalent widths in 
such regions of short wavelength (actually, our measurements are practically 
confined to lines of $\lambda \ga 5200$~\AA; see Figure~4c and 4f), because  
they are crowded with strong spectral lines (note that the locations of large 
discrepancy are mostly accompanied by strong line absorptions; cf. Figure~11a).
} 

There are, however, some cases where the spectrum disentangling did not turn out 
successful. A typical example is seen in the $\sim$~6280--6320~\AA\ region 
(cf. Figure~11d) where prominently sharp spike-like discrepancies are observed 
between $r^{\rm obs}$ and $r^{\rm rec}$. Such features are also recognized 
(though less significantly) in the neighborhood of H$\alpha$ (see Figure~11f).   
This is due to a number of conspicuous telluric lines, the wavelengths of which remain 
unchanged irrespective of orbital phases unlike stellar lines. Figure~11d$'$ indicates that our 
$r^{\rm rec}$ fails to reproduce not only the telluric lines but also the stellar line profiles.
Accordingly, we can learn that the decomposed spectra processed from the data including 
many strong lines of telluric origin should not be used. This is the reason why we prepared the 
special disentangled spectra of 6290--6310\AA\ region (for the analysis of [O~{\sc i}] 6300 line)
by removing the telluric line in advance from the original $r^{\rm obs}$ (cf. Section~3).
Figure~11e$'$ shows that the disentangling in the relevant region has been made successful
by this preprocessing.

\subsection{Line-widths check for the secondary}

It is important for successful spectrum disentangling to assign sufficiently accurate 
radial velocities for both the primary and secondary components at each phase 
[$v_{\rm p}^{\rm local}(\phi)$, $v_{\rm s}^{\rm local}(\phi)$].
While there must be no problem for the primary star because we could directly
measure $v_{\rm p}^{\rm local}(\phi)$ on each spectrum, there may be some concern
for the secondary, for which we calculated $v_{\rm s}^{\rm local}(\phi)$
according to Equation~(4). Although we confirmed from theoretical simulations 
carried out for the analysis of O~{\sc i}~7771--5 triplet (cf. Section 5.1) that 
$v_{\rm p}^{\rm local}$ and $v_{\rm s}^{\rm local}$ resulting from Equations (3) 
and (4) yielded quite satisfactory consistency between the theoretical and 
observed primary+secondary composite spectra, additional check would be desirable. 

If inadequate $v_{\rm s}^{\rm local}(\phi)$ values were chosen which do not 
correspond to the actual shifts of spectral lines, the lines profiles in the
resulting disentangled spectra of the secondary would unnaturally show excessive
broadening. Then, we can make use of the results of $v_{\rm e}\sin i$ (projected
rotational velocity) derived as a by-product of spectral fitting analysis
(cf. Section~5.1), where the disentangled spectra were employed in most cases, 
while the composite double-line spectrum was exceptionally used only for the 
analysis of O~{\sc i} 7771--5 triplet. That is, if a consistency of $v_{\rm e}\sin i$ 
could be confirmed between the former and the latter, we may state that our 
choice of $v_{\rm s}^{\rm local}(\phi)$ is justified.

The $v_{\rm e}\sin i$ solutions based on the disentangled secondary spectra 
turned out to be 31.8, 32.3, 32.3, 33.2, 34.1, 33.7, 34.1, ad 33.6~km~s$^{-1}$
from the 5378.5--5382~\AA, 6080--6089~\AA,  6101--6105~\AA, 
6157--6164~\AA, 6297--6303~\AA, 6361--6365~\AA, 6704--6711~\AA, and 
6751--6760~\AA\ regions, respectively, which yield 33.1~km~s$^{-1}$
as the mean (standard deviation is 0.9~km~s$^{-1}$). Meanwhile, 
the $v_{\rm e}\sin i$ of the secondary component we obtained from 
the 7768--7786~\AA\ region analysis of the composite primary+secondary spectrum 
was 33.7~km~s$^{-1}$, which reasonably matches the mean value mentioned above.  
Accordingly, the agreement between these two $v_{\rm e}\sin i$ values derived 
from the spectra of different nature may be counted as another evidence for 
the adequacy of spectrum disentangling.     

\subsection{Comparison of the primary with four Hyades giants}

Although the most desirable and straightforward way to verify the validity of the 
disentangled spectrum is to compare it with the ``true'' spectrum obtained 
directly by spatially resolved observations, such a genuine spectrum of Capella's 
each component is unfortunately not available to us.     
However, we know that Capella (belonging to Hyades moving group) and Hyades stars 
are considered to be siblings, and the Hyades cluster contains four evolved red giants:
HD~27371 ($\gamma$~Tau), HD~27697 ($\delta$~Tau), HD~28305 ($\epsilon$~Tau), 
and HD~28307 ($\theta^{1}$~Tau). These four Hyades giants (G8~{\sc iii}--K0~{\sc iii}) 
and the primary of Capella (G8~{\sc iii}) are quite similar to each other, which are
considered to be red-clump giants in the core He-burning stage as seen from 
the luminosities of these four Hyades stars ($\log L/L_{\odot} \sim$1.9--2.0; 
cf. Takeda et al. 2008) and that of the primary ($\log L/L_{\odot} = 1.90$).
Accordingly, we may use these four stars as a ``proxy'' for the primary component 
of Capella. That is, if the equivalent widths measured on the decomposed spectrum 
of Capella's primary and those of Hyades giants are confirmed to be  
consistent with each other, this may be regarded as an indirect proof that our 
disentangling procedure was adequately done (or at least without any serious problem).
 
The atmospheric parameters ($T_{\rm eff}$, $\log g$, $v_{\rm t}$, and [Fe/H]) of 
these Hyades giants, which were spectroscopically determined by Takeda et al. (2008) 
in the same manner as adopted in this study, are graphically compared with
those of the primary (cf. Section 4) in Figures~12a--12d, where we can see that
they are mostly in accord with each other. Especially, the fact that almost the same
metallicity results ([Fe/H]~$\sim +0.1$) were derived for all these five stars 
is meaningful, since the metallicity is expected to be practically equal in this case
(unlike other parameters for which small star-to-star difference may be possible).
Considering that these parameters were determined from equivalent widths ($W_{\lambda}$) 
of Fe~{\sc i} and Fe~{\sc ii} lines, we can imagine the similarity of $W_{\lambda}$ 
for all these stars. Actually, this consistency can be corroborated as demonstrated 
in Figure~12e--12h, where our $W_{\lambda}$ values of Fe lines for the primary star
measured on the disentangled spectrum (cf. ``felines.dat'') are compared with
those of HD~23371, 27697, 28305, and 28307 (which were taken from tableE2 of 
Takeda et al. 2008). Consequently, we may conclude also in this respect that 
the true spectrum has been adequately reproduced by our decomposed spectrum 
to a practically sufficient precision.

\onecolumn

\clearpage

\begin{table}[h]
\small
\caption{Basic information of observed data and relevant radial velocities.}
\begin{center}
\begin{tabular}{ccccccccc
}\hline\hline                 
Obs. date   &    HJD      & $\phi$ &  $\delta V^{\rm hel}$ & $V_{\rm p}^{\rm local}$ & 
$V_{\rm p}^{\rm hel}$ & $V_{\rm s}^{\rm hel}$ & $V_{\rm s}^{\rm local}$ & Site/Instrument\\
(1) & (2) & (3) & (4) & (5) & (6) & (7) & (8) & (9) \\
\hline
2010/04/29 & 2455315.931 & 0.8634 & $-$19.07 & +64.9 & +45.83 & +10.17 & +29.24 & Okayama/HIDES\\
\hline
2015/11/30 & 2457357.170 & 0.4866 &  +6.10 &  $-$7.2 &  $-$1.10 & +57.10 & +51.00 & Gunma/GAOES\\
2015/12/18 & 2457375.282 & 0.6607 &  $-$2.54 & +17.4 & +14.86 & +41.14 & +43.68 & Gunma/GAOES\\
2016/01/14 & 2457402.165 & 0.9191 & $-$14.93 & +63.2 & +48.27 &  +7.73 & +22.66 & Gunma/GAOES\\
2016/02/02 & 2457421.082 & 0.1010 & $-$21.80 & +69.1 & +47.30 &  +8.70 & +30.50 & Gunma/GAOES\\
2016/02/15 & 2457434.002 & 0.2252 & $-$25.17 & +57.3 & +32.13 & +23.87 & +49.04 & Gunma/GAOES\\
2016/03/01 & 2457449.050 & 0.3698 & $-$27.32 & +36.0 &  +8.68 & +47.32 & +74.64 & Gunma/GAOES\\
2016/03/15 & 2457463.010 & 0.5040 & $-$27.73 & +29.2 &  +1.47 & +54.53 & +82.26 & Gunma/GAOES\\
2016/03/16 & 2457463.959 & 0.5132 & $-$27.75 & +29.2 &  +1.45 & +54.55 & +82.30 & Gunma/GAOES\\
2016/03/22 & 2457469.995 & 0.5712 & $-$27.31 & +30.3 &  +2.99 & +53.01 & +80.32 & Gunma/GAOES\\
2016/03/24 & 2457471.983 & 0.5903 & $-$27.13 & +32.7 &  +5.57 & +50.43 & +77.56 & Gunma/GAOES\\
2016/03/25 & 2457472.978 & 0.5999 & $-$27.03 & +33.9 &  +6.87 & +49.13 & +76.16 & Gunma/GAOES\\
\hline
\end{tabular}
\end{center}
(1) Observation date (yyyy/mm/dd; in UT). (2) Heliocentric Julian day. (3) Orbital phase, 
which was calculated with HJD($\phi = 0$) = 2450857.21 + 104.02173$E$ ($E$: integer). 
(4) Heliocentric correction (km~s$^{-1}$) computed by the ``rvcorrect'' task of IRAF. 
(5) Actually observed local topocentric radial velocity (km~s$^{-1}$) for the primary.
(6) Heliocentric radial velocity (km~s$^{-1}$) for the primary evaluated from
$\delta V^{\rm hel}$ and $V_{\rm p}^{\rm local}$.
(7) Heliocentric radial velocity (km~s$^{-1}$) for the secondary evaluated as 
$56 - V_{\rm p}^{\rm hel}$ (km~s$^{-1}$) based on Equation~(1) and Equation~(2). 
(8) Local topocentric radial velocity (km~s$^{-1}$) for the secondary
computed from $\delta V^{\rm hel}$ and $V_{\rm s}^{\rm hel}$.
(9) Observing site and used instrument. 
\end{table}

\clearpage

\begin{table}[h]
\small
\caption{Evaluated equivalent widths and non-LTE corrections for important lines.}
\begin{center}
\begin{tabular}{cccccl
}\hline\hline                 
 Object  & $W_{\lambda}$ & $A^{\rm L}$ & $A^{\rm N}$ & $\Delta^{\rm N}$ & Remark \\ 
  (1)  & (2) & (3)  & (4)   & (5)  & (6) \\
\hline
\multicolumn{5}{c}{[C~{\sc i} 5380]} & \\
primary          &   16.8 &  8.501 &  8.480 & $-$0.021  & \\
secondary        &   45.6 &  8.744 &  8.701 & $-$0.043  & \\
Sun              &   19.9 &  8.680 &  8.672 & $-$0.008  & \\
\hline                                    
\multicolumn{5}{c}{[Li~{\sc i} 6104]} & \\
primary          &$\cdots$&$\cdots$&$\cdots$&$\cdots$ & undetectable\\
secondary        &   20.9 &  3.338 &  3.393 & +0.055  & \\
Sun              &$\cdots$&$\cdots$&$\cdots$&$\cdots$ & undetectable\\
\hline                                    
\multicolumn{5}{c}{[Na~{\sc i} 6161]} & \\
primary          &  124.5 &  6.878 &  6.720 & $-$0.158  & \\
secondary        &   48.6 &  6.161 &  6.090 & $-$0.071  & \\
Sun              &   59.0 &  6.364 &  6.306 & $-$0.058  & \\
\hline                                    
\multicolumn{5}{c}{[Zn~{\sc i} 6362]} & \\
primary          &   41.0 &  4.826 &  4.769 & $-$0.057  & \\
secondary        &   45.0 &  4.681 &  4.651 & $-$0.030  & \\
Sun              &   19.4 &  4.500 &  4.495 & $-$0.005  & \\
\hline                                    
\multicolumn{5}{c}{[Li~{\sc i} 6708]} & \\
primary          &   12.4 &  0.782 &  0.988 & +0.206  & \\
secondary        &  194.4 &  3.235 &  3.176 & $-$0.059  & \\
Sun              &    3.1 &  1.029 &  1.096 & +0.067  & \\
\hline                                    
\multicolumn{5}{c}{[S~{\sc i} 6757]} & \\
primary          &   22.3 &  7.268 &  7.240 & $-$0.028  & \\
secondary        &   48.0 &  7.274 &  7.232 & $-$0.042  & \\
Sun              &   20.1 &  7.207 &  7.202 & $-$0.005  & \\
\hline
\multicolumn{5}{c}{[O~{\sc i} 7774]} & \\
primary          &   53.1 &  9.139 &  8.960 & $-$0.179  & double-line spectrum \\
secondary        &  144.6 &  9.515 &  9.088 & $-$0.427  & double-line spectrum \\
Sun              &   64.4 &  8.982 &  8.879 & $-$0.103  & \\
\hline
\end{tabular}
\end{center}
(1) Object. (2) Equivalent width (m\AA). See Table~3 of Paper~I for the data of 
the relevant spectral lines. (3) LTE abundance. (4) Non-LTE abundance.
(5) Non-LTE correction ($\equiv A^{\rm N} - A^{\rm L}$). (6) Specific remark.
Note that $A^{\rm L}$ and $A^{\rm N}$ (logarithmic number abundances) are 
expressed in the usual normalization of $A({\rm H}) = 12.00$.
\end{table}

\clearpage

\begin{table}[h]
\scriptsize
\caption{Results of relative abundances for the primary and secondary components.}
\begin{center}
\begin{tabular}{ccccccl
}\hline\hline
$Z$& Species & [X/H]$_{\rm p}$ & [X/H]$_{\rm s}$ & [X$_{\rm s}$/X$_{\rm p}$] & 
$\lambda_{1}$--$\lambda_{2}$ or $N_{\rm p}$/$N_{\rm s}$ & Remark \\
(1) & (2) & (3) & (4) & (5) & (6) & (7) \\
\hline
\multicolumn{7}{c}{[Spectrum synthesis analysis]}\\
 3 &  Li  & $\cdots$ &  3.393   & $\cdots$  &  6101--6105 & $A$(Li) instead of [Li/H], NLTE\\
 3 &  Li  &  0.988  &  3.176   &  +2.188   &  6704--6711 & $A$(Li) instead of [Li/H], NLTE\\
 6 &  C   & $-$0.192  & +0.029   & +0.221   &  5378.5--5382 & NLTE \\
 8 &  O   & +0.006  & $\cdots$  & $\cdots$  &  6297--6303 & \\
 8 &  O   & $-$0.323  & $\cdots$  & $\cdots$  &  6361--6365 & \\
 8 &  O   & +0.081  & +0.209   & +0.128   &  7768--7786 & NLTE\\
11 &  Na  & +0.414  & $-$0.216   & $-$0.630   &  6157--6164 & NLTE\\
14 &  Si  & +0.169  & $-$0.061   & $-$0.230   &  6080--6089 & \\
14 &  Si  & +0.175  & +0.019   & $-$0.156   &  6297--6303 & \\
16 &  S   & +0.038  & +0.030   & $-$0.008   &  6751--6760 & NLTE\\
20 &  Ca  & +0.038  & $-$0.170   & $-$0.208   &  6101--6105 & \\
20 &  Ca  & +0.174  & $-$0.251   & $-$0.425   &  6157--6164 & \\
21 &  Sc  & $-$0.011  & $\cdots$  & $\cdots$  &  6297--6303 & \\
22 &  Ti  & +0.376  & $-$0.184   & $-$0.560   &  5378.5--5382 & \\
22 &  Ti  & $-$0.042  & $-$0.163   & $-$0.121   &  6080--6089 & \\
23 &  V   & +0.167  & $-$0.062   & $-$0.229   &  6080--6089 & \\
24 &  Cr  & +0.256  & $\cdots$  & $\cdots$  &  6361--6365 & \\
26 &  Fe  & +0.214  & $-$0.384   & $-$0.598   &  5378.5--5382 & \\
26 &  Fe  & +0.029  & $-$0.255   & $-$0.284   &  6080--6089 & \\
26 &  Fe  & +0.007  & $-$0.010   & $-$0.017   &  6101--6105 & \\
26 &  Fe  & $-$0.026  & $-$0.220   & $-$0.194   &  6157--6164 & \\
26 &  Fe  & $-$0.123  & $-$0.127   & $-$0.004   &  6297--6303 & \\
26 &  Fe  & +0.201  & $-$0.094   & $-$0.295   &  6361--6365 & \\
26 &  Fe  & +0.057  & $-$0.201   & $-$0.258   &  6704--6711 & \\
26 &  Fe  & +0.086  & $-$0.109   & $-$0.195   &  6751--6760 & \\
26 &  Fe  & $-$0.045  & $-$0.593   & $-$0.548   &  7768--7786 & \\
27 &  Co  & +0.146  & $-$0.599   & $-$0.745   &  5378.5--5382 & \\
27 &  Co  & +0.210  & +0.074   & $-$0.136   &  6080--6089 & \\
28 &  Ni  & +0.008  & $-$0.306   & $-$0.314   &  6080--6089 & \\
28 &  Ni  & +0.086  & $-$0.525   & $-$0.611   &  6157--6164 & \\
28 &  Ni  & +0.027  & $-$0.325   & $-$0.352   &  6361--6365 & \\
30 &  Zn  & +0.274  & +0.156   & $-$0.118   &  6361--6365 & NLTE\\
\hline
\multicolumn{7}{c}{[Equivalent width analysis]}\\
14 &  Si~{\sc i}  &$+0.20(\pm 0.12)$ & $\cdots$ & $\cdots$ & 17/-- & \\
20 &  Ca~{\sc i}  &$+0.16(\pm 0.07)$ & $\cdots$ & $\cdots$ &  7/-- & \\
21 &  Sc~{\sc ii} &$+0.02(\pm 0.10)$ & $\cdots$ & $\cdots$ &  7/-- & \\
22 &  Ti~{\sc i}  &$+0.07(\pm 0.08)$ & $\cdots$ & $\cdots$ & 29/-- & \\
23 &  V~{\sc i}   &$+0.15(\pm 0.11)$ & $\cdots$ & $\cdots$ &  8/-- & \\
24 &  Cr~{\sc i}  &$+0.17(\pm 0.09)$ & $\cdots$ & $\cdots$ & 14/-- & \\
26 &  Fe~{\sc i}  &$+0.10(\pm 0.05)$ & $-0.08(\pm 0.08)$& $-0.18(\pm 0.09)$ & 147/52 & Param. determ. by-product\\
26 &  Fe~{\sc ii}  &$+0.10(\pm 0.06)$ & $-0.08(\pm 0.12)$& $-0.18(\pm 0.13)$ & 16/6 & Param. determ. by-product\\
27 &  Co~{\sc i}  &$+0.08(\pm 0.16)$ & $\cdots$ & $\cdots$ &  8/-- & \\
28 &  Ni~{\sc i}  &$+0.05(\pm 0.10)$ & $-0.10(\pm 0.19)$ & $-0.15(\pm 0.13)$ & 36/10 & \\
\hline
\end{tabular}
\end{center}
(1) Atomic number. (2) Element or species. (3) Relative abundance of the primary to the Sun
$[\equiv A_{\rm p}({\rm X}) - A_{\odot}({\rm X})]$.
(4) Relative abundance of the secondary to the Sun
$[\equiv A_{\rm s}({\rm X}) - A_{\odot}({\rm X})]$.
(5) Relative abundance of the secondary to the primary
 ($[\equiv A_{\rm s}({\rm X}) - A_{\rm p}({\rm X})]$).
(6) Range (in \AA) of spectrum fitting (for spectrum synthesis analysis)
or the number of used lines for the primary/secondary (for equivalent width analysis).
(7) Specific remark.
\end{table}

\clearpage

\begin{table}[h]
\small
\caption{Published results for the atmospheric parameters, metallicity, and Li abundance of Capella.
}
\begin{center}
\begin{tabular}{cccccc
}\hline\hline
Component & $T_{\rm eff}$ & $\log g$ & $v_{\rm t}$ & [Fe/H] & $A$(Li) \\
 (1) & (2) & (3) & (4) & (5) & (6) \\
\hline
\multicolumn{6}{c}{[McWilliam (1990)]} \\
Combined$^{a}$  & 5270 & 3.05 & $2.0^{b}$ & $-0.37^{c}$ & $\cdots$ \\
\hline
\multicolumn{6}{c}{[Pilachowski \& Sowell (1992)]} \\
Primary   & 4800 & 2.6 & 2.0 & $+0.08^{d}$ & 0.8 \\
Secondary & 5550 & 2.8 & 1.0 & $-0.08^{d}$ & 3.0 \\
\hline
\multicolumn{6}{c}{[Randich et al. (1994)]} \\
Primary   & 4900 & 2.6 & $\cdots$ & $-0.4$ & 0.6 \\
Secondary & 5550 & 2.8 & $\cdots$ & $0.0$ & 3.1 \\
\hline
\multicolumn{6}{c}{[Fuhrmann (2011)]} \\
Primary   & 4951 & 2.68 & 1.4 & $+0.05^{e}$ & $\cdots$ \\
Secondary & 5672 & 2.94 & 1.8 & $+0.05^{e}$ & $\cdots$ \\
\hline
\multicolumn{6}{c}{[Torres et al. (2015)]} \\
Primary   & 4980 & 2.69 & 1.48 & $-0.03^{f}$ & 1.08 \\
Secondary & 5750 & 2.94 & 1.55 & $-0.06^{f}$ & 3.28 \\
\hline
\multicolumn{6}{c}{[This study]} \\
Primary   & 4943 & 2.52 & 1.47& $+0.10^{g}$ & $0.99^{h}$ \\
Secondary & 5694 & 2.88 & 2.29& $-0.08^{g}$ & $3.18^{h}$ \\
\hline
\end{tabular}
\end{center}
\scriptsize
(1) Relevant component. (2) Effective temperature (K). (3) Logarithmic
surface gravity (cm~s$^{-2}$/dex). (4) Microturbulent velocity (km~s$^{-1}$).
(5) Fe abundance relative to the Sun (dex). (6) Lithium abundance.\\ \\
$^{a}$Presumably, the apparent line-strengths of the primary component were 
measured in the double-lined spectrum and simply used as they are.\\
$^{b}$Assumed $v_{\rm t}$.\\
$^{c}$Reduced to $-0.20$ if the recent solar abundance of 7.50 is used (cf. Torres to al. 2009).\\
$^{d}$[Ca/H] instead of [Fe/H].\\ 
$^{e}$[Fe/H]$_{\rm p}$=[Fe/H]$_{\rm s}$ is assumed.\\
$^{f}$Result from Fe~{\sc i} lines.\\
$^{g}$Result from the equivalent widths of Fe~{\sc i} lines as a by-product of atmospheric parameter determination.\\
$^{h}$Non-LTE result from the Li~{\sc i} 6708 line.\\
\end{table}

\clearpage

\begin{figure}
\epsscale{.65}
\plotone{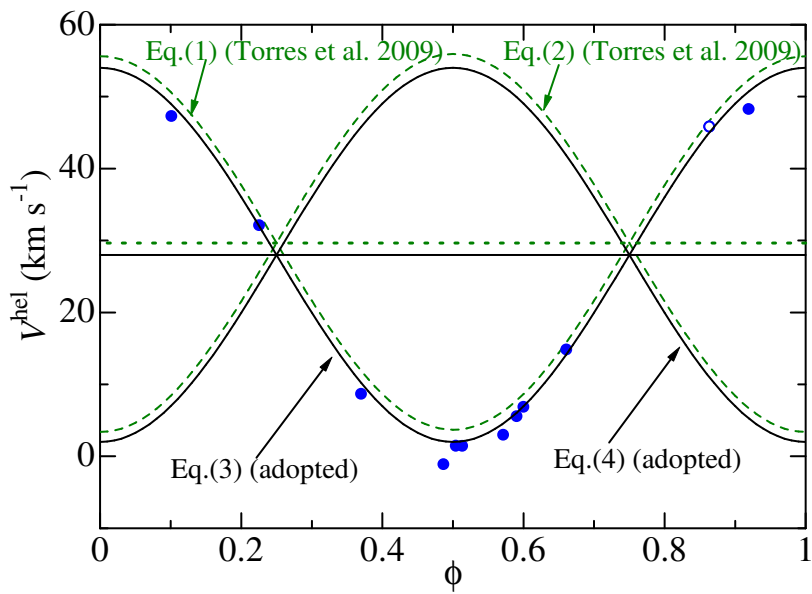}
\caption{
Symbols: heliocentric radial velocities of the primary 
 ($V_{\rm p}^{\rm hel}$; cf. Table~1) corresponding to each of 
the 12 spectra (filled circles: GAOES data, open circle: OAO data), 
plotted against the orbital phase ($\phi$). 
Dashed lines: Torres et al.'s (2009) $V_{\rm p}^{\rm hel}(\phi)$ and 
$V_{\rm s}^{\rm hel}(\phi)$ relations [Equations (1) and (2)]. 
Solid lines: Finally adopted $V_{\rm p}^{\rm hel}(\phi)$ and 
$V_{\rm s}^{\rm hel}(\phi)$ relations [Equations (3) and (4)], in which 
the system velocity was slightly adjusted so as to fit better with 
the observed $V_{\rm p}^{\rm hel}$ (symbols).
}
\end{figure}

\clearpage

\begin{figure}
\epsscale{.65}
\plotone{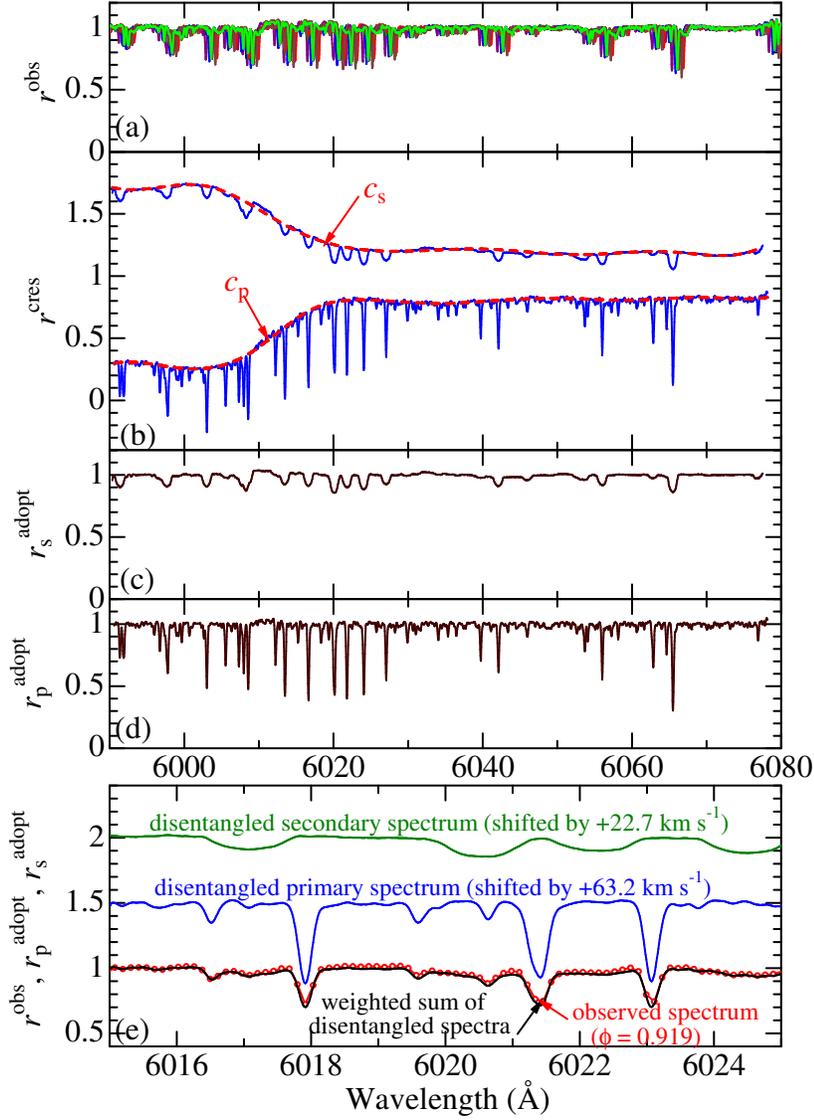}
\caption{
(a) Overplotted are the original 11 GAOES spectra ($r^{\rm obs}$; continuum-normalized 
residual flux in the 5900--6080~\AA\ region) used for spectrum disentangling, where 
only the $\phi = 0.504$ spectrum is highlighted in light green.
(b) Solid lines: disentangled spectra of the primary ($r_{\rm p}^{\rm cres}$) 
and the secondary ($r_{\rm s}^{\rm cres}$), which were obtained as the direct 
output of CRES program. Dashed lines: apparent continuum ($c_{\rm p}$ and $c_{\rm s}$), 
which was defined as the envelope of $r_{\rm p}^{\rm cres}$ and $r_{\rm s}^{\rm cres}$, respectively.
(c) Finally adopted disentangled spectrum of the secondary ($r_{\rm s}^{\rm adopt}$).
(d) Finally adopted disentangled spectrum of the primary ($r_{\rm p}^{\rm adopt}$).
(e) Demonstration of how the adopted disentangle spectra of the primary 
($r_{\rm p}^{\rm adopt}$; top) and the secondary ($r_{\rm s}^{\rm adopt}$; middle) 
reproduce the original spectrum ($r^{\rm obs}$; open triangles; bottom) for 
the case of $\phi = 0.919$, when combined by applying the appropriate shifts 
corresponding to the relevant $V_{\rm p}^{\rm local}$ and $V_{\rm s}^{\rm local}$ 
along with the adopted flux ratio (the resulting spectrum is depicted by the 
solid line at the bottom). Each spectrum in top, middle, and bottom is shifted 
by 0.5 relative to the adjacent one. 
}
\end{figure}

\clearpage

\begin{figure}
\epsscale{.70}
\plotone{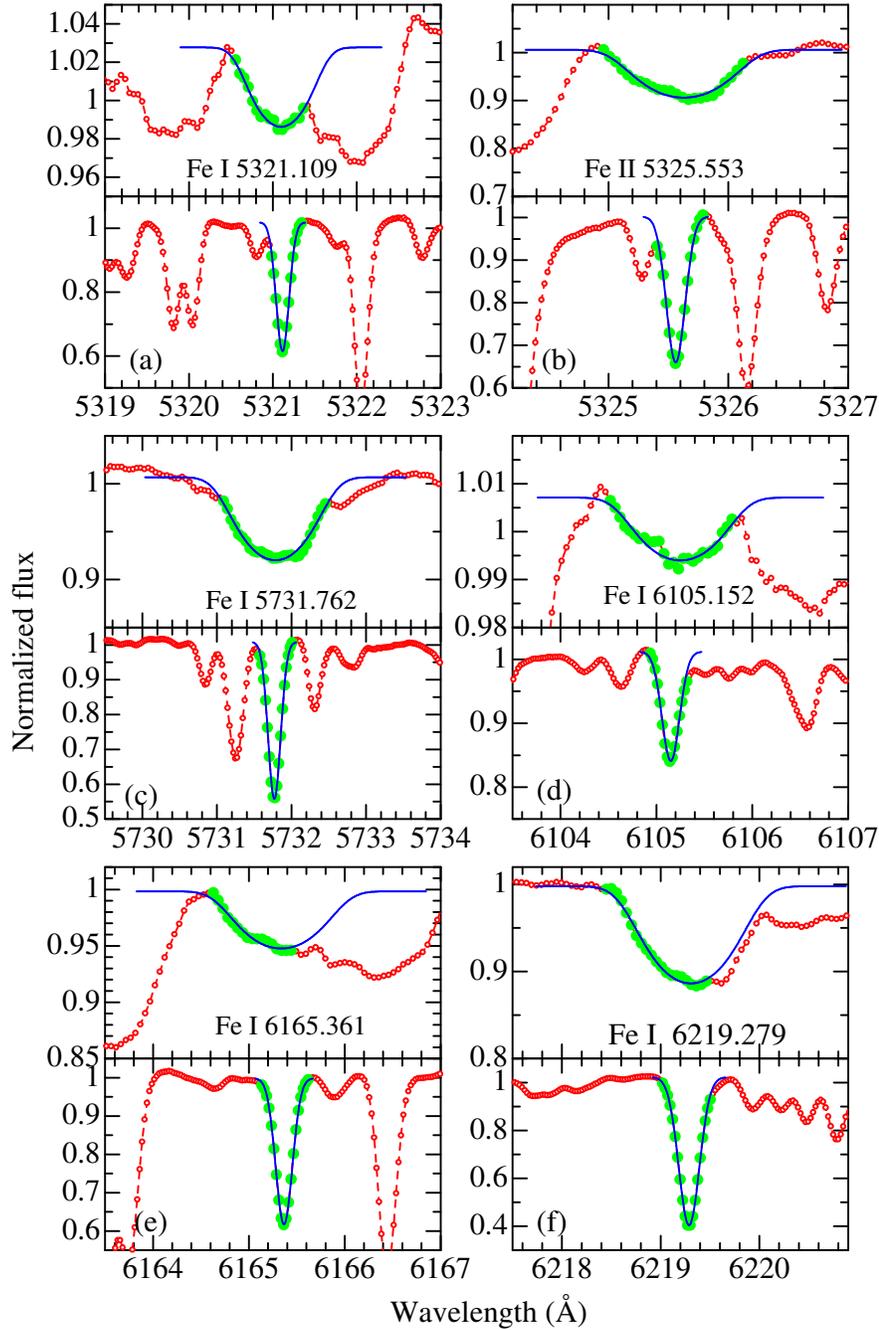}
\caption{
Examples of how we measured the equivalent widths for the secondary (upper panel)
and the primary (lower panel). The observed data are shown by symbols, while
the fitted function (rotational+Gaussian for the secondary and Gaussian for
the primary) is depicted by solid lines. The part of the observed profile,
which was used to determine the fitting function, are indicated by larger symbols 
in light green. (a) Fe~{\sc i} 5321.109, (b) Fe~{\sc ii} 5325.553, (c) Fe~{\sc i} 5731.762, 
(d) Fe~{\sc i} 6105.152, (e) Fe~{\sc i} 6165.361, and (f) Fe~{\sc i} 6219.279.
}
\end{figure}

\clearpage

\begin{figure}
\epsscale{.70}
\plotone{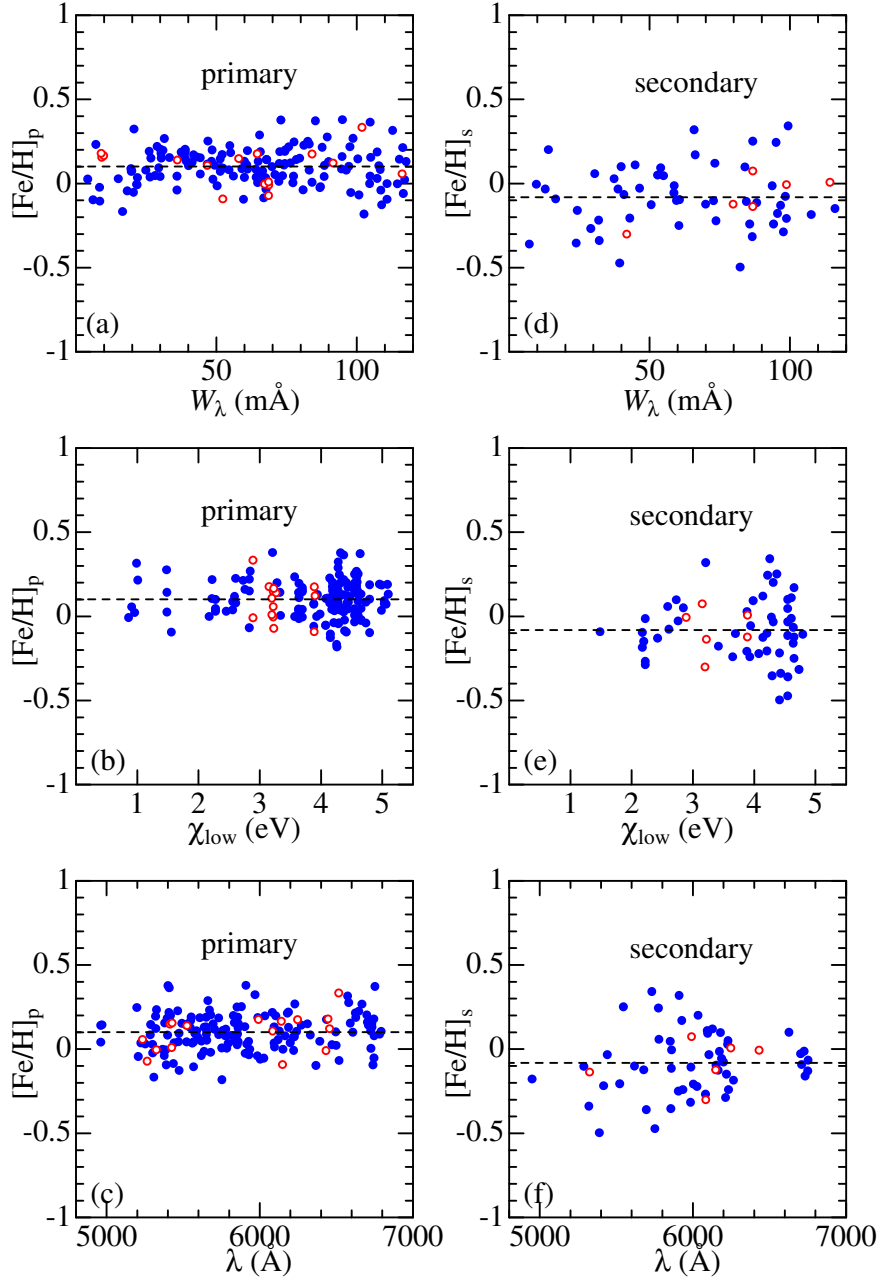}
\caption{
[Fe/H] vs. $W_{\lambda}$ (equivalent width) relation (upper panels: a, d),
[Fe/H] vs. $\chi_{\rm low}$ (lower excitation potential) relation (middle panels: b, e), and 
[Fe/H] vs. $\lambda$ (wavelength) relation (bottom panels: c, f) 
corresponding to the finally established atmospheric parameters of 
$T_{\rm eff}$, $\log g$, and $v_{\rm t}$. The left-side panels (a, b, c) are for 
the primary while the right-side panels (d, e, f) are for the secondary.
The filled and open symbols correspond to Fe~{\sc i} and Fe~{\sc ii} lines, 
respectively. The positions of the mean [Fe/H] (+0.10 for the primary and $-0.08$ 
for the secondary) are indicated by the horizontal dashed lines.
}
\end{figure}

\clearpage

\begin{figure}
\epsscale{.90}
\plotone{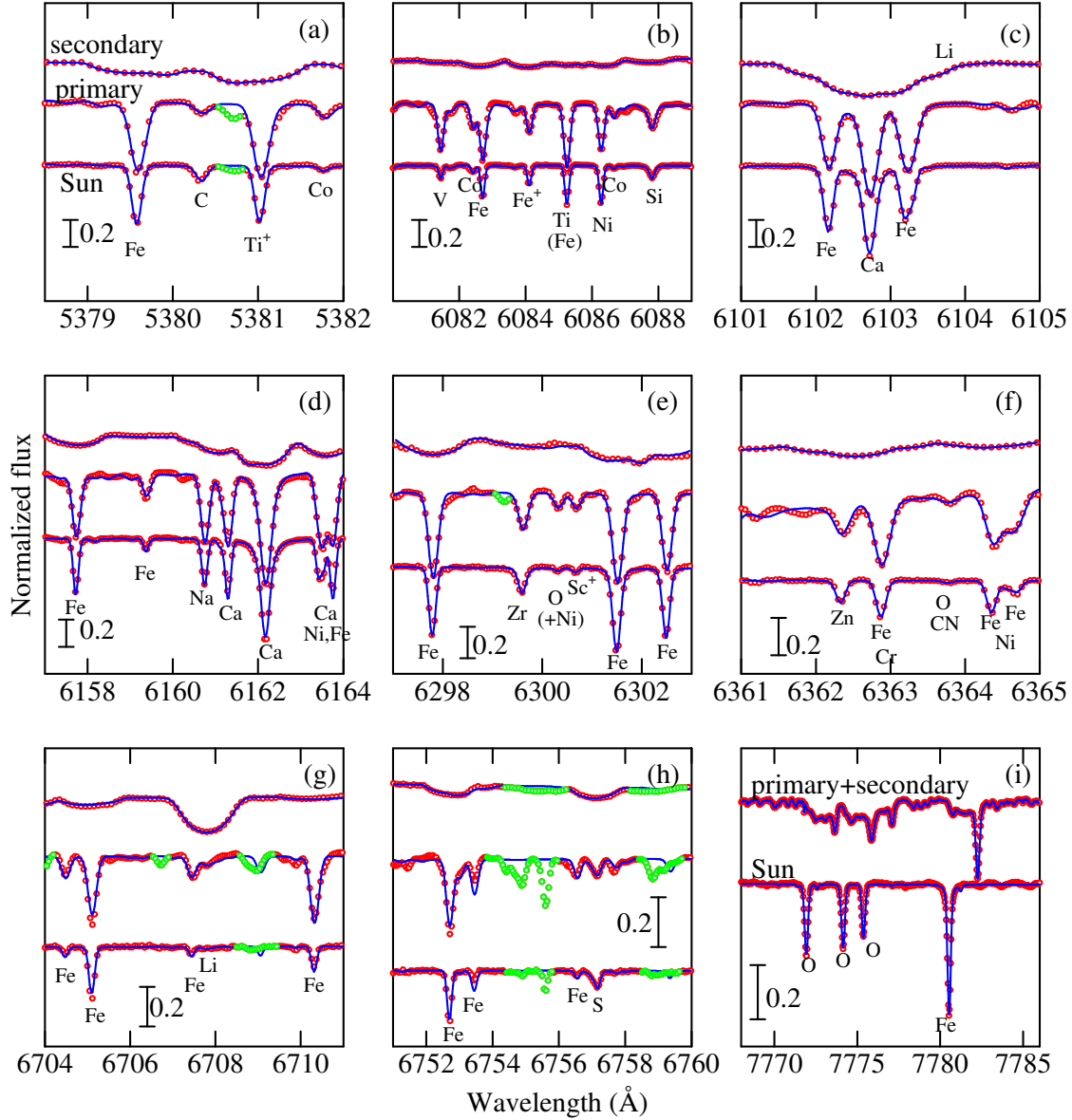}
\caption{
Synthetic spectrum fitting analysis carried out for 9 wavelength regions
for abundance determinations. The observed spectra are plotted in red 
symbols (where the masked regions discarded in judging the goodness of fit 
are colored in light-green) while the best-fit theoretical spectra are 
shown in blue solid lines. 
(a) 5378.5--5382~\AA\ region (for C, Ti, Fe, Co), 
(b) 6080--6089~\AA\ region (for Si, Ti, V, Fe, Co, Ni), 
(c) 6101--6105~\AA\ region (for Li, Ca, Fe),
(d) 6157--6164~\AA\ region (for Na, Ca, Fe, Ni),
(e) 6297--6303~\AA\ region (for O, Si, Sc, Fe),
(f) 6361--6365~\AA\ region (for O, Cr, Fe, Ni, Zn),
(g) 6704--6711~\AA\ region (for Li, Fe)
(h) 6751--6760~\AA\ region (for S, Fe), and
(i) 7768--7786~\AA\ region (for O, Fe).
Note that telluric lines in the 6297--6303~$\rm\AA$ region (panel e)
had been removed before the disentangling procedure was applied (cf. Section~3).
Similarly, the broad Ca~{\sc i} autoionization feature in the 6361--6365~$\rm\AA$ 
region (panel f) was smoothed out by redrawing the continuum level.
}
\end{figure}

\clearpage

\begin{figure}
\epsscale{.60}
\plotone{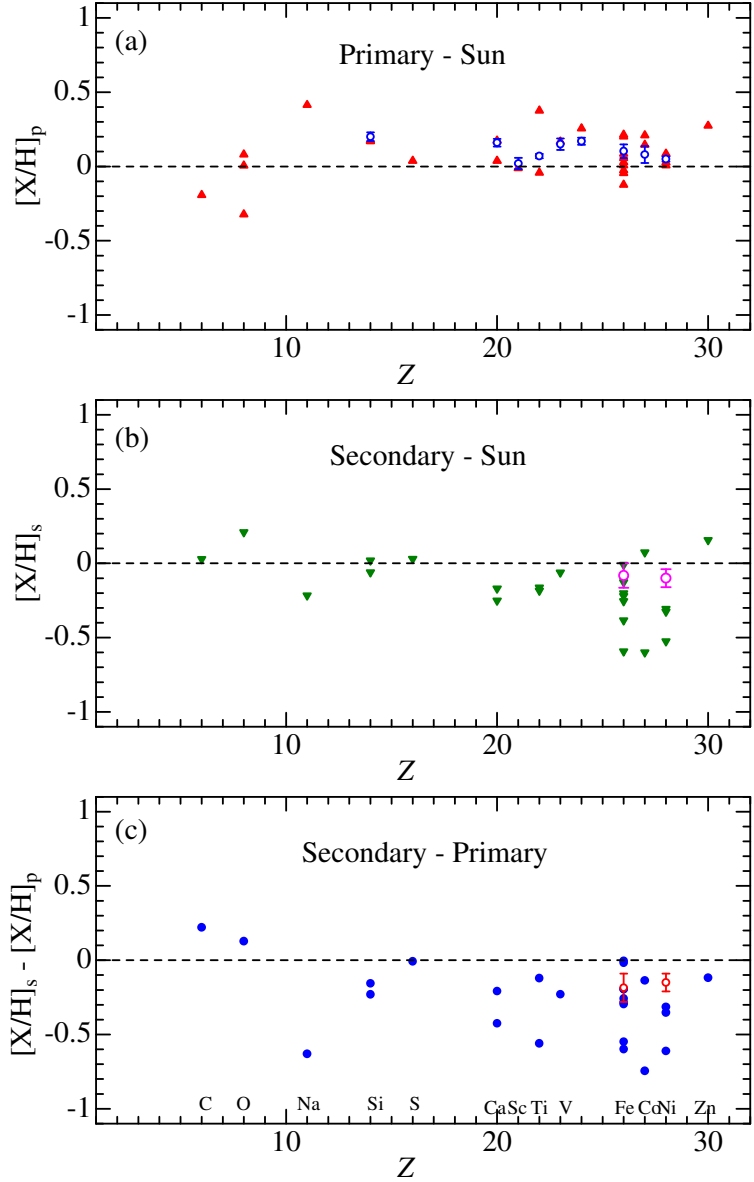}
\caption{
Differential abundances of various elements plotted against the atomic number ($Z$).
The results derived from spectrum synthesis analysis are shown by filled 
symbols, while those from equivalent width analysis are by open symbols
(attached error bar denotes the mean error $\sigma/\sqrt{N}$ as in Section~4). 
(a) Abundances of the primary relative to the Sun.
(b) Abundances of the secondary relative to the Sun.
(c) Abundances of the secondary relative to the primary.
Note that the results for Li are not shown here.
}
\end{figure}

\clearpage

\begin{figure}
\epsscale{.80}
\plotone{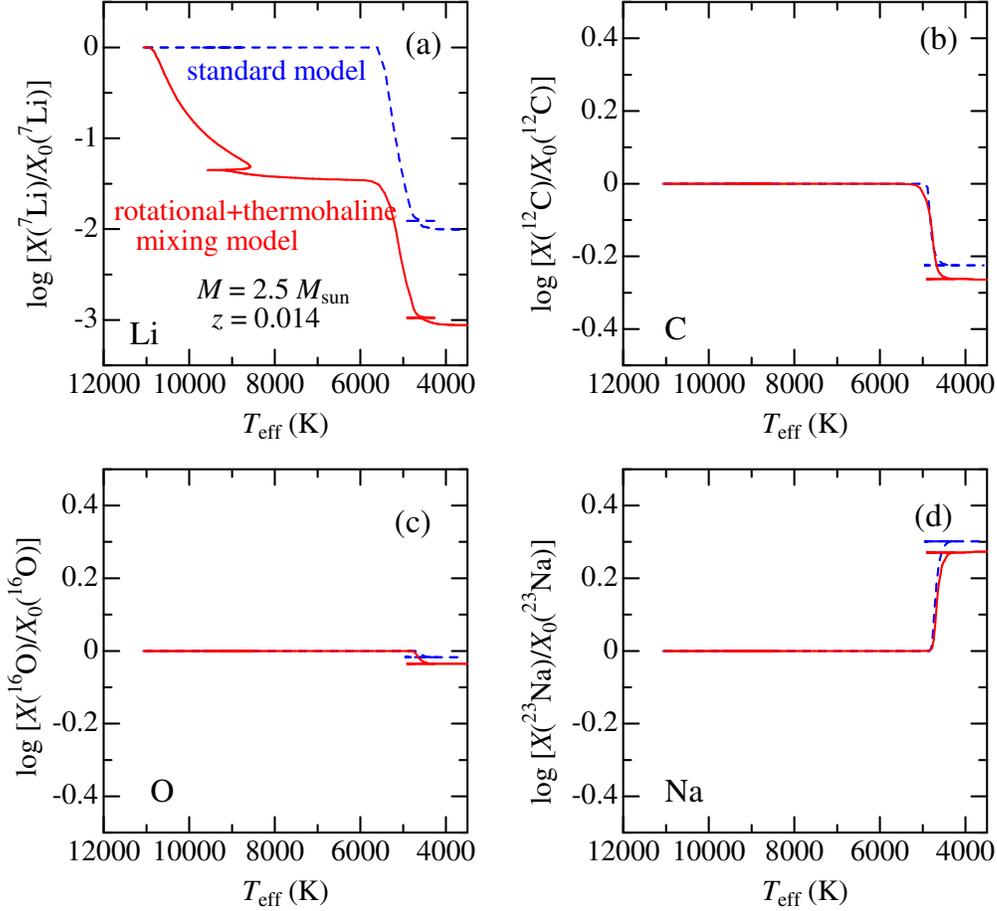}
\caption
{Run of the logarithmic surface abundances relative to the initial compositions 
($\log [X/X_{0}]$) for selected light elements are plotted against $T_{\rm eff}$, 
which were theoretically simulated along the evolutionary sequence for 
the $M = 2.5 M_{\odot}$ solar-metallicity model by Lagarde et al. (2012), 
where two different treatments of envelope mixing were adopted: (i) standard mixing
model (dashed line) and (ii) non-canonical model including rotational+thermohaline 
mixing with the initial rotation rate of $0.45\times$critical velocity (solid line). 
(a) $^{7}$Li, (b) $^{12}$C, (c) $^{16}$O, and (d) $^{23}$Na. 
}
\end{figure}

\clearpage

\begin{figure}
\epsscale{.60}
\plotone{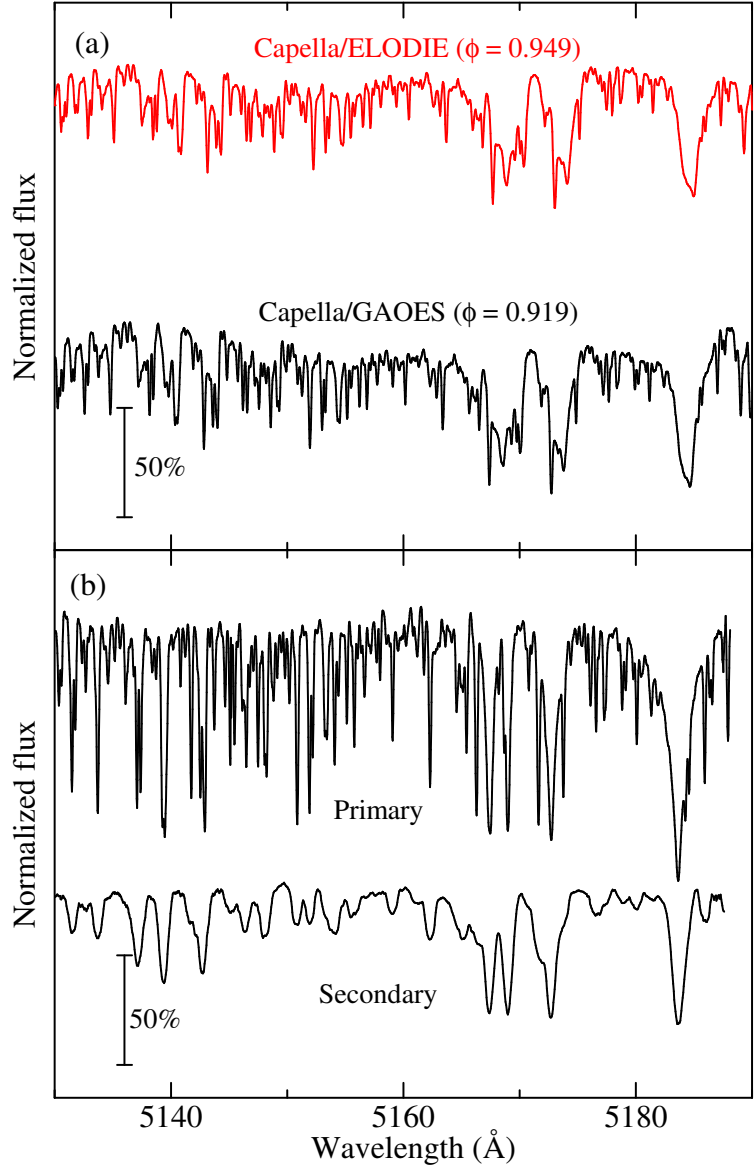}
\caption
{
(a) Selected 5130--5190~\AA\ portion of ELODIE spectrum of Capella (observed on 
2003 March 25 corresponding to $\phi = 0.949$; elodie\_20030325\_0023.fits) compared
with that of our GAOES spectrum (observed on 2016 January 14 corresponding to $\phi = 0.919$).  
(b) Our disentangled spectra for the primary and the secondary in the
5130--5190~\AA\ region, which are to be compared with Torres et al.'s (2015) Figure~2. 
}
\end{figure}

\clearpage

\begin{figure}
\epsscale{.70}
\plotone{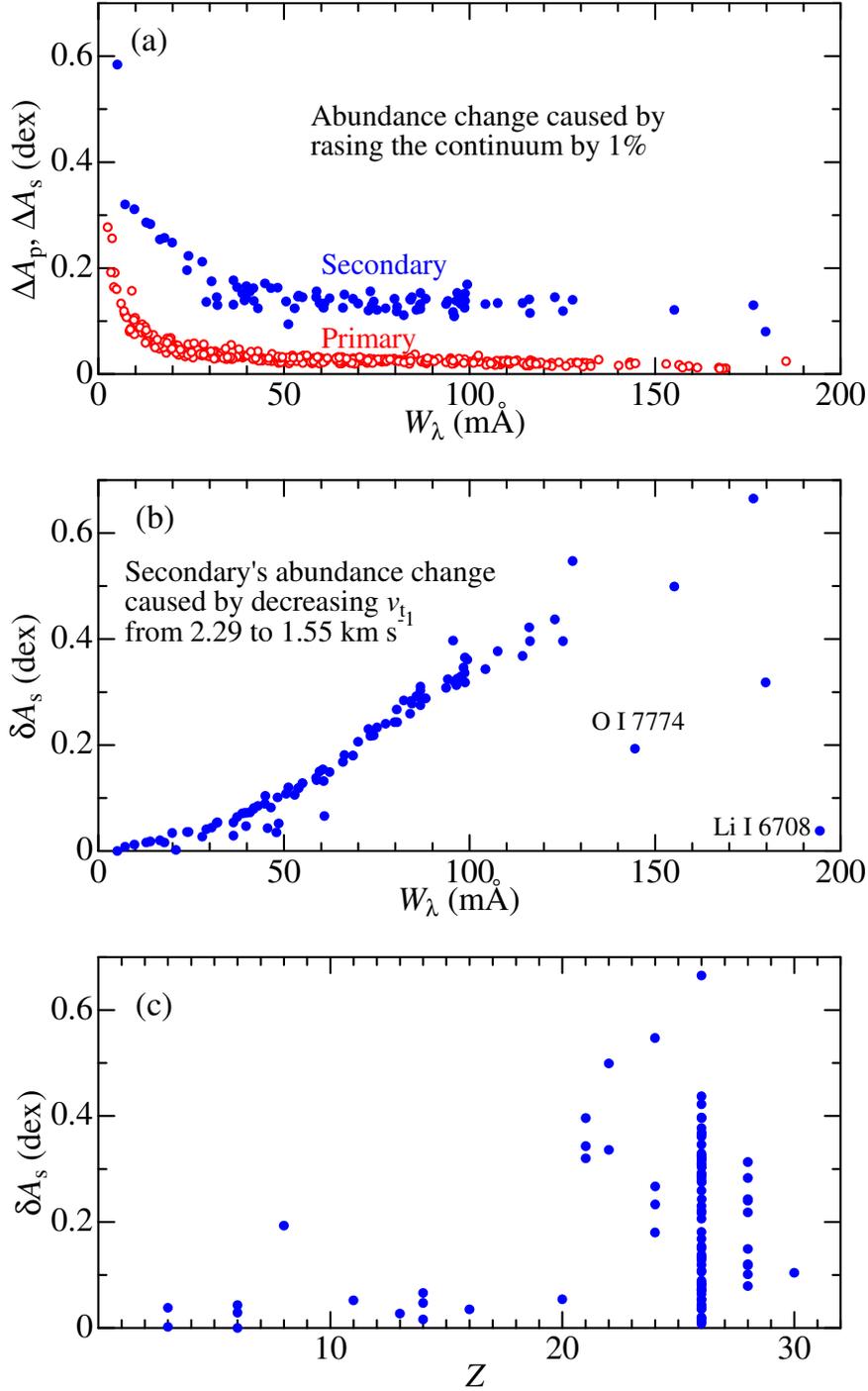}
\caption
{
(a) Abundance variations ($\Delta A$) due to the changes in $W_{\lambda}$ 
caused by raising the continuum level by 1\% (which were estimated
as decribed in Section~6.4.2) plotted against $W_{\lambda}$.
Open circles are for the primary and filled circles are for the secondary.
(b) Abundance changes for the secondary ($\delta A_{\rm s}$) caused by
using the microturbulence of $v_{\rm t}$ = 1.55~km~s$^{-1}$ (the value used 
by Torres et al. 2015) instead of our adopted value of $v_{\rm t}$ = 
2.29~km~s$^{-1}$, plotted against $W_{\lambda}$.
(c) Abundance changes for the secondary ($\delta A_{\rm s}$) in response to
the reduction of $v_{\rm t}$ as in (b), plotted against the atomic number ($Z$).
}
\end{figure}

\clearpage

\begin{figure}
\epsscale{.70}
\plotone{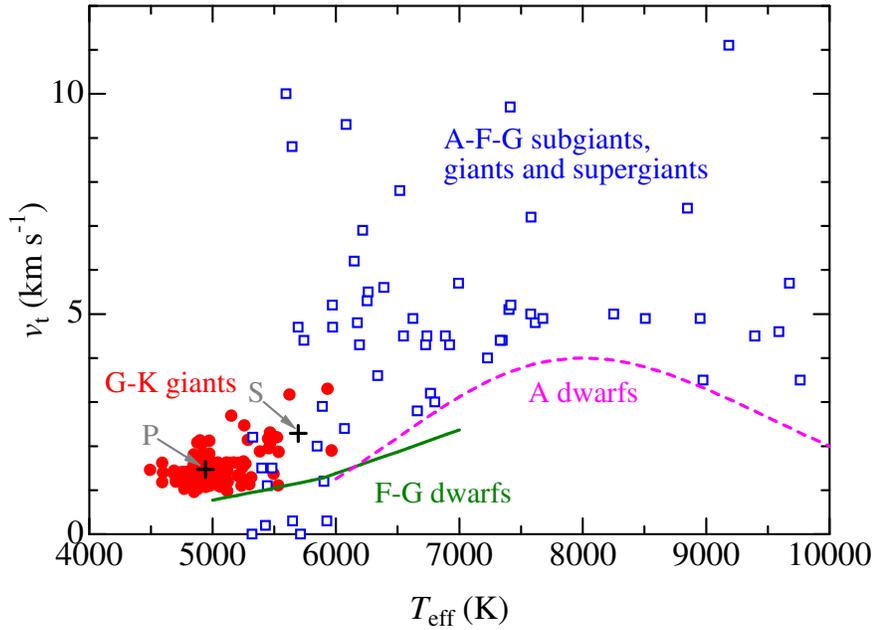}
\caption
{
Microturbulence ($v_{\rm t}$) values of evolved stars plotted against $T_{\rm eff}$.
Filled circles: G and early K giants (taken from Takeda et al. 2008; cf. Fig.~1d therein).
Open squares: A--F--G subgiats, giants, and supergiants (taken from Takeda et al.
2018; abundance-based microturbulence denoted as $\xi_{\rm a}$; cf. Fig.~6b therein).
Depicted in lines are the mean relations for dwarf stars (solid line $\cdots$
FG dwarfs, dashed line $\cdots$ A dwarfs; see Takeda et al. 2018 for more details).
Our $v_{\rm t}$ values for the primary and the secondary are indicated by two crosses. 
}
\end{figure}

\clearpage

\begin{figure}
\epsscale{.60}
\plotone{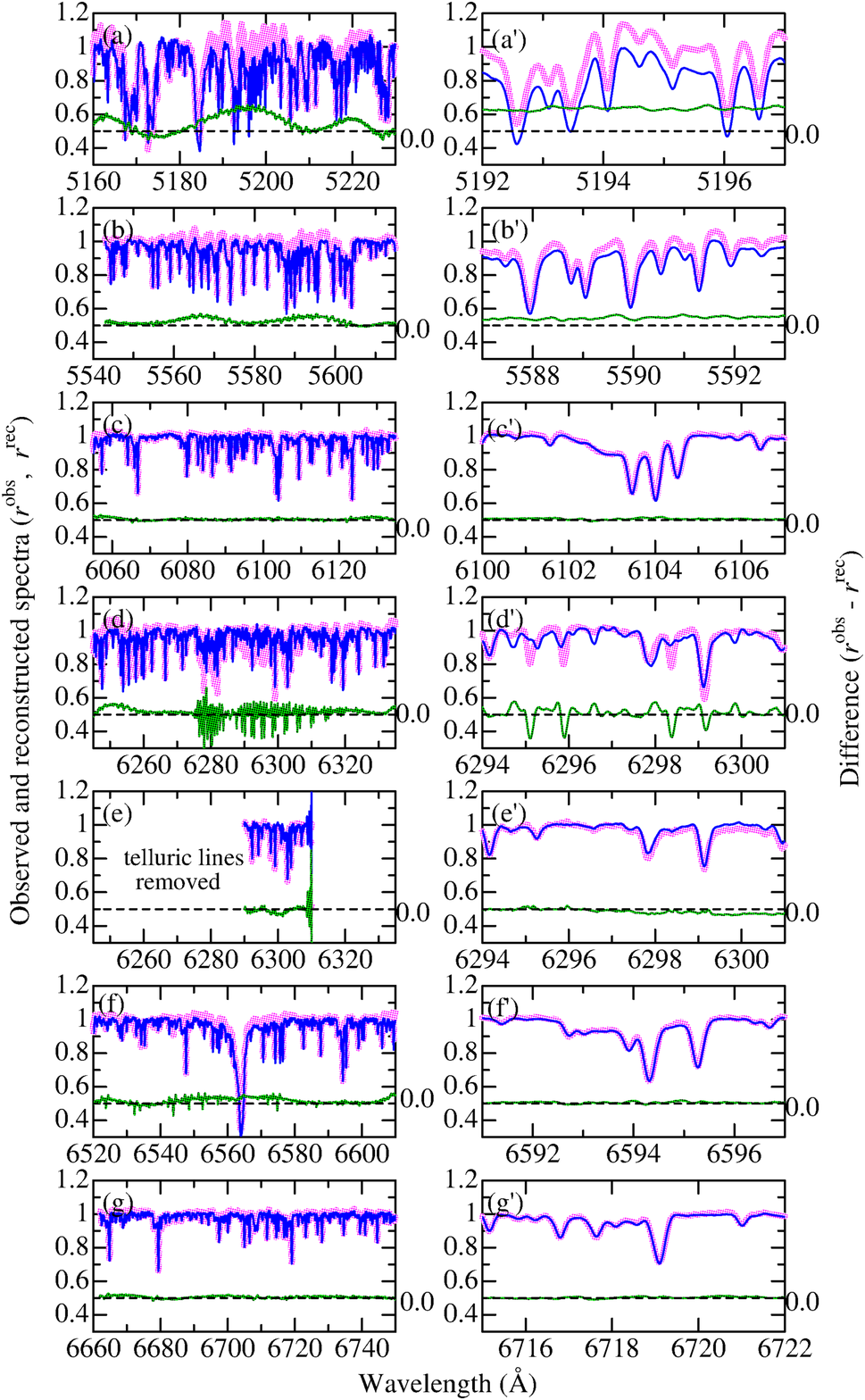}
\caption
{
Comparison of the reconstructed primary+secondary spectrum  ($r^{\rm rec}$; 
blue lines), which was simulated from the disentangled spectra 
($r_{\rm p}^{\rm adopt}$ and $r_{\rm s}^{\rm adopt}$ being shifted with the 
relevant local radial velocities) by Equation~(A1), with the actually observed 
double-line spectrum ($r^{\rm obs}$ at $\phi = 0.919$; pink symbols). 
The green line shows the difference between these two spectra 
($r^{\rm obs} - r^{\rm rec}$), where a vertical offset of
0.5 has been applied (the zero level is indicated by a dashed line and the 
origin is marked in the right-hand axis). The left-hand panels show the
spectra of six echelle orders [(a) 5160--5230~\AA, (b) 5540--5615~\AA, 
(c) 6055--6135~\AA, (d) 6245--6335~\AA, 
(e) 6290--6310~\AA (similar to panel (d) 
but telluric lines were removed in advance of disentangling),
(f) 6520--6610~\AA,
and (g) 6660--6750~\AA], while the magnified spectra in selected narrow 
wavelength regions are depicted in the corresponding right-hand panels
(a$'$)--(g$'$). 
}
\end{figure}

\clearpage

\begin{figure}
\epsscale{.70}
\plotone{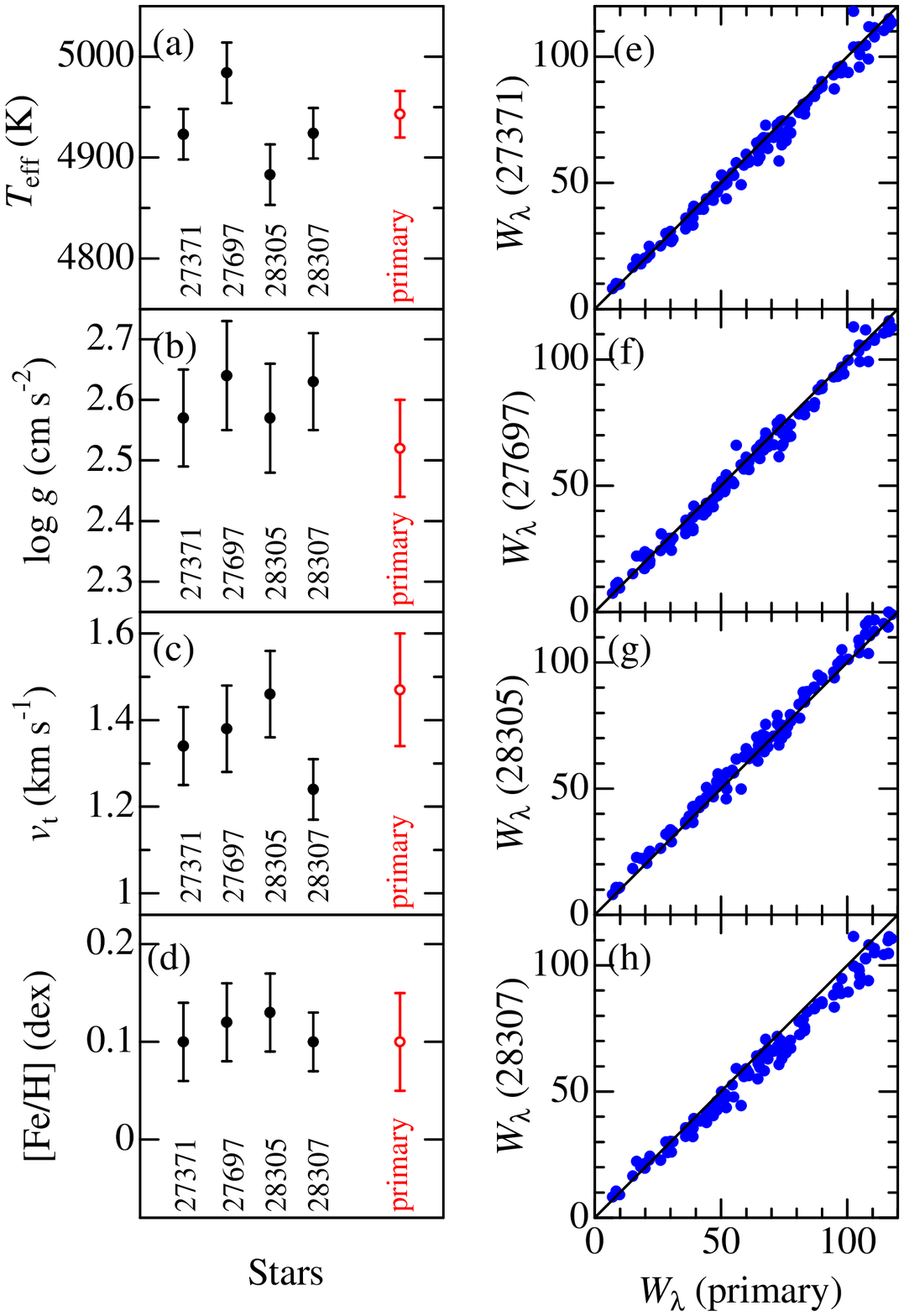}
\caption
{
The left-hand panels (a)--(d) illustrate the mutual comparison of the atmospheric 
parameters ($T_{\rm eff}$, $\log g$, $v_{\rm t}$, and [Fe/H]) of four Hyades 
giants (HD~27371, 27697, 28305, and 28307; filled circles) and the primary 
of Capella (open circles). As in Section~4. the attached error bars 
in panels (a)--(c) indicate the statistical errors, while those in panel (d) 
are the mean errors. 
The right-hand panels (e)--(h) show graphical comparisons of the equivalent widths 
of Fe~{\sc i} and Fe~{\sc ii} lines for the primary of Capella 
(which we measured  in this study based on the disentangled spectra) with those 
of four Hyades giants (taken from Takeda et al. 2008) having similar atmospheric 
parameters to the primary: (e) HD~27371, (f) HD~27697, (g) HD~28305, and (h) HD~28307.
}
\end{figure}

\clearpage






\clearpage






\begin{thebibliography}{}
\bibitem[]{}
  Asplund, M., Grevesse, N., Sauval, A. J., \& Scott, P. 2009, ARA\&A, 47, 481
\bibitem[]{}
  Boesgaard, A. M., Budge, K. G. 1988, ApJ, 332, 410 
\bibitem[]{}
  Eggen, O. J. 1960, MNRAS, 120, 540
\bibitem[]{}
  Eggen, O. J. 1972, PASP, 84, 406
\bibitem[]{}
  Fuhrmann, K. 2011, ApJ, 742, 42
\bibitem[]{}
  Gray, D. F. 1978, Solar Phys., 59, 193
\bibitem[]{}
  Gray, D. F. 1982, ApJ, 262, 682
\bibitem[]{}
  Gray, D. F. 1988, Lectures on Spectral-Line Analysis: F, G, and K stars
  (Arva, Ontario, The Publisher)
\bibitem[]{}
  Gray, D. F. 2005, The Observation and Analysis of Stellar Photospheres, 3rd ed.
  (Cambridge, Cambridge University Press)
\bibitem[]{}
  Hadrava, P. 1995, A\&AS, 114, 393
\bibitem[]{}
  Hensberge, H., Iliji\'{c}, S., \& Torres, K. B. V. 2008, A\&A, 482, 1031
\bibitem[]{}
  Iliji\'{c}, S. 2004, in Spectroscopically and Spatially Resolving the Components
  of Close Binary Stars, ASP Conf. Ser. Vol. 318, ed. R. W. Hilditch, H. Hensberge,
  \& K. Pavlovski (San Francisco: Astronomical Society of the Pacific), 107 
\bibitem[]{}
  Kurucz, R. L. 1993, Kurucz CD-ROM, No. 13 (Harvard-Smithsonian Center
  for Astrophysics)
\bibitem[]{}
  Kurucz, R.~L., \& Bell B. 1995, Kurucz CD-ROM, No. 23 
  (Cambridge: Smithsonian Astrophysical Observatory) 
\bibitem[]{}
  Kurucz, R. L., Furenlid, I., Brault, J., \& Testerman, L. 1984,  
  Solar Flux Atlas from 296 to 1300 nm
  (Sunspot, New Mexico: National Solar Observatory)
\bibitem[]{}
  Lagarde, N., Decressin, T., Charbonnel, C., Eggenberger, P., Ekstr\"{o}m, S.,
  \& Palacios, A. 2012, A\&A, 543, A108
\bibitem[]{}
  McWilliam, A. 1990, ApJS, 74, 1075
\bibitem[]{}
  Moultaka, J., Ilovaisky, S. A., Prugniel, P., \& Soubiran, C. 2004, PASP, 116, 693
\bibitem[]{}
  Pilachowski, C. A., \& Sowell, J. R. 1992, AJ, 103, 1668
\bibitem[]{}
  Randich, S., Giampapa, M. S., \& Pallavicini, R. 1994, A\&A, 283, 893
\bibitem[]{}
  Sanad, M. R. 2013, Ap\&SS, 344, 389
\bibitem[]{}
  Simon, K. P., \& Sturm, E. 1994, A\&A, 281, 286
\bibitem[]{}
  Takeda, Y. 1995a, PASJ, 47, 287
\bibitem[]{}
  Takeda, Y. 1995b, PASJ, 47, 463
\bibitem[]{}
  Takeda, Y. 2008, in The Metal-Rich Universe, eds. G. Israelian \& G. Meynet,
  (Cambridge: Cambridge University Press), 308
\bibitem[]{}
  Takeda, Y., et al. 2005a, PASJ, 57, 13
\bibitem[]{}
  Takeda, Y., Jeong, G., \& Han, I. 2018, PASJ, 70, 8
\bibitem[]{}
  Takeda, Y., Honda, S., Ohnishi, T., Ohkubo, M, Hirata, R., \& Sadakane, K. 2013,
  PASJ, 65, 53
\bibitem[]{}
  Takeda, Y., Ohkubo, M., \& Sadakane, K. 2002, PASJ, 54, 451
\bibitem[]{}
  Takeda, Y., Ohkubo, M., Sato, B., Kambe, E., \& Sadakane, K. 2005b, PASJ, 57, 27
  [Erratum: PASJ, 57, 415]
\bibitem[]{}
  Takeda, Y., Sato, B., Kambe, E., Izumiura, H., Masuda, S., \& Ando, H. 2005c, PASJ, 57, 109
\bibitem[]{}
  Takeda, Y., Sato, B., \& Murata, D. 2008, PASJ, 60, 781 
\bibitem[]{}
  Takeda, Y., Sato, B., Omiya, M., \& Harakawa, H. 2015, PASJ, 67, 24
\bibitem[]{}
  Takeda, Y., \& Tajitsu, A. 2017, PASJ, 69, 74 (Paper~I)
\bibitem[]{}
  Torres, G., Claret, A., \& Young, P. A. 2009, ApJ, 700, 1349
\bibitem[]{}
  Torres, G., Claret, A., Pavlovski, K., \& Dotter, A. 2015, ApJ, 807, 26
\bibitem[]{}
  van Bueren, H. G. 1952, Bull. Astron. Inst. Neth. Supl. Ser. 11, 385
\bibitem[]{}
  Zhao, J., Zhao, G., \& Chen, Y. 2009, ApJ, 692, L113

\end{thebibliography}
\end{document}